\documentclass[nofootinbib,showkeys,floatfix,onecolumn,times]{revtex4-2}
\usepackage{bm,morefloats}
\usepackage{dcolumn}
\usepackage[latin1]{inputenc}
\usepackage[T2A,T1]{fontenc} 
\usepackage{fonttable}
\usepackage{graphicx}
\usepackage{balance}
\usepackage[latin1]{inputenc}
\usepackage[T1]{fontenc}
        \usepackage{cellspace}
\usepackage{latexsym,color}
\usepackage{centernot}
\usepackage{mathtools}
\usepackage{stmaryrd}
\usepackage{amssymb}
\usepackage{natbib}
\usepackage{amsmath,amssymb,amsthm,mathrsfs,amsfonts,dsfont}

\usepackage{color}

\usepackage{rotating}
\usepackage[dvipsnames]{xcolor}
\definecolor{LinkBlue}{RGB}{6,69,173}
\definecolor{DarkBlue}{RGB}{11,0,128}
\definecolor{red}{rgb}{1,0.,0.}
\usepackage[colorlinks=true,linkcolor=LinkBlue,urlcolor=magenta,
	citecolor=green,hyperfootnotes=true,urlcolor=red]{hyperref}

\usepackage{enumitem}
\count\footins = 1000

\newcommand{\prdd}{Phys. Rev. \textbf{D}}

\begin{document}
\count\footins = 1000

\title{Horizon replicas in black hole shadows}
\author{D. Pugliese$^{1}$  and H. Quevedo$^{2}$}
\email{daniela.pugliese@physics.slu.cz,quevedo@nucleares.unam.mx}
%

%
\affiliation{ $^1$ Research Centre for Theoretical Physics and Astrophysics,
Institute of Physics,
	Silesian University in Opava,
	Bezru\v{c}ovo n\'{a}m\v{e}st\'{i} 13, CZ-74601 Opava, Czech Republic
	\\     $^{2}$
	Instituto de Ciencias Nucleares, Universidad Nacional Aut\'onoma de M\'exico,   AP 70543, M\'exico, DF 04510, Mexico \\
		Dipartimento di Fisica and ICRA, Universit\`a di Roma ``La Sapienza", I-00185 Roma, Italy }

\begin{abstract}
Recently,  new exploratory channels have opened up for the physics of highly compact objects, such as gravitational waves and black hole shadows. Moreover,  more precise analysis and observations are now possible in the physics of accretion around compact objects. These advancements provide in particular  an  unprecedented insight into the physics near the  horizons of a black hole. In this work we focus on the shadow boundary of a Kerr black hole,  introducing   observables related to special null orbits, called horizons replicas,  solutions of the shadow edge equations  which are related to particular photon orbits, defined by constraints   on their impact parameter,   carrying  information about the angular momentum of the central spinning object.
These orbits are related to  particular regions on the shadow  boundary  and might be used to determine the spin of the black hole.
The results provide the  conditions by which horizon replicas are imprinted in the black hole  shadow profile, in dependence on the  black hole dimensionless spin and observational angle,    providing eventually
new templates for the future  observations.
\end{abstract}
%
\date{\today}

\maketitle

\def\be{\begin{equation}}
\def\ee{\end{equation}}
\def\bea{\begin{eqnarray}}
\def\eea{\end{eqnarray}}
\newcommand{\tb}[1]{\textbf{\texttt{#1}}}
\newcommand{\actaa}{Acta Astronomica}
\newcommand{\laa}{\mathcal{L}}
\newcommand{\al}{\mathcal{A}}
\newcommand{\ba}{\mathcal{B}}
\newcommand{\MBs}{\mathbf{MBs}}
\newcommand{\Na}{\mathcal{N}}
\newcommand{\La}{\mathcal{L}}
\newcommand{\Em}{\mathcal{E}}

\newcommand{\mso}{\mathrm{mso}}
\newcommand{\mbo}{\mathrm{mbo}}

\newcommand{\rtb}[1]{\textcolor[rgb]{1.00,0.00,0.00}{\tb{#1}}}
\newcommand{\gtb}[1]{\textcolor[rgb]{0.17,0.72,0.40}{\textbf{#1}}}
\newcommand{\ptb}[1]{\textcolor[rgb]{0.77,0.04,0.95}{\tb{#1}}}
\newcommand{\btb}[1]{\textcolor[rgb]{0.00,0.00,1.00}{\textbf{#1}}}
\newcommand{\otb}[1]{\textcolor[rgb]{1.00,0.50,0.25}{\tb{#1}}}
\newcommand{\non}[1]{{\LARGE{\not}}{#1}}

\newcommand{\cc}{\mathrm{C}}

\newcommand{\il}{~}
\newcommand{\la}{\mathcal{A}}
  \newcommand{\Qa}{\mathcal{Q}}
\newcommand{\Sa}{\mathcal{\mathbf{S}}}
\newcommand{\Ta}{{\mbox{\scriptsize  \textbf{\textsf{T}}}}}
\newcommand{\Ca}{\mathcal{\mathbf{C}}}
\section{Introduction}
In this work, we study some aspects of
 black hole (\textbf{BH}) shadow boundary  (edge/critical curve) related to   special null orbits called  horizons replicas,   characterized by impact parameter   (specific angular  momentum)
$\ell=\ell_H^\pm(a)$ equal in magnitude to the angular momentum of the outer and inner \textbf{BH} horizons, respectively ($a$ is the \textbf{BH} dimensionless spin). The analysis introduces  new observables related to  replicas and discussing the relation with the
  \textbf{BH} spin  and its
  inclination angle with the respect to an observer at infinity.

The   observation of the close proximity of the  \textbf{BH} horizon (\textbf{BH} shadow)   located at the center of the giant elliptical galaxy
\textbf{M87} by the \textbf{E}vent \textbf{H}orizon \textbf{T}elescope (\textbf{EHT}) Collaboration\footnote{https://eventhorizontelescope.org.} constitutes a consistent  advance in observational astronomy, allowing    a tight interplay between  theoretical  analysis and the  observations-- see \cite{EHT1,EHT2,EHT2c,EHT2d,EHT2e,EHT2f,
EHT21a,EHT21b} (and  \cite{Medeiros} for a recent  sharper image of the \textbf{M87} \textbf{BH}, created with PRIMO algorithm).
 More recently the \textbf{EHT} Collaboration has been   able to observe  also the shadow, as   asymmetric rings of synchrotron emissions, cast by
the super-massive \textbf{BH}  (\textbf{SMBH})    Sagittarius \textbf{A*} (\textbf{SgrA*}), located at  the center of our galaxy \cite{EHT22a,EHT22b,EHT22c,
EHT22d,EHT22e,EHT22f}.

 The \textbf{EHT} observations, provide in particular  a new insight in the   physical phenomena in the   close   proximity of the  \textbf{BH}  event horizon.
On the other hand,  \textbf{BHs} shadows have been studied extensively analytically--see for example \cite{Synge,Luminet,Bardeen1973,Chandra}--where
we should note that  main
 notion of the   \textbf{BH}   shadow refers mostly to   the shadow edge
  (the shadow luminous boundary), rather than  the shadow itself,
being  the dark  region bounded by
 the luminous edge,  constituted by  unstable  (spherical and circular) photon orbits-- \cite{Bardeen1973,Synge,Luminet,Chandra,Hioki,Johannsen,Ghasemi-Li-Bambi,Rezzolla}.
 In many analysis the   \textbf{BH}, assumed  embedded in a luminous  environment,  is   considered  located before the screen and the distant observer.
 The  photons,  moving  towards the  \textbf{BH},  could  be  trapped in
unstable circular orbits,  and  can  reach  the far away  observer after perturbation. Photons could escape to infinity and  detected by the distant observer     in light of sight, or  they can  be captured by the  central  \textbf{BH}  generating  the  \textbf{BH} shadow as the central dark region.
The \textbf{BH} shadow  profile (luminous edge)  can be  also  affected by the  matter environment  and  could  depend also on properties   of the region of the  light
distribution  and its  source\footnote{In these conditions,  the photons could   interact with  the accreting  plasma  modifying both   photons distribution and the intensity--see  for example
\cite{FMA00,T04,BD05,BL06,
BN06,HCS07,Doku}.}.

However, in all cases, the    shadow profile morphology reflects     the \textbf{BH} parameters, which   can be estimated by the analysis of observable   related to the topological and morphological properties of the shadow boundary curve, constructed for example, using special points on the
shadow boundary in the celestial coordinates $(\alpha,\beta)$ (see also \cite{Tamburini,Hioki,Johannsen,Ghasemi-Li-Bambi,Rezzolla}.  These    observables for a Kerr \textbf{BH}  carry information on the central spinning attractor  and the photon orbits constituting the luminous texture of the boundary.
In this work,  we assume that photons are generated by a distant source  at infinity and are distributed uniformly in \emph{all}
directions, observers are  also located at infinity, and that the Kerr \textbf{BH} has a well defined and fixed\footnote{A further interesting aspect would be to investigate \textbf{BH} shadows following a spin variation (precession) process.}
 inclination angle $\theta$, which is defined by the
angle between the \textbf{BH}  rotation axis and the observer line of
sight.

In general, the \textbf{BH} shadow profile  is found considering photons having  \emph{all} the possible values of the impact parameters $\ell$, therefore solving the equations  for the shadow profile, system   $(\mathfrak{R})$, for \emph{all} values of $\ell$.
In our  analysis, we solve  $(\mathfrak{R})$ in dependence on the \textbf{BH} dimensionless spin  $a\in [0,1]$  and of the
  observational $\theta$ angle,  constraining  assuming $\ell=\pm\ell_H^\pm(a)$, for  co-rotating and counter-rotating motion (determined by the $\ell$ parameter).
By solving  $(\mathfrak{R})$, within the constraints $\ell=\pm\ell_H^\pm(a)$,   we examine the parts of the luminous edges (shadows profiles) associated to the  null geodesics at fixed constraints, mapping in this way the \textbf{BH} luminous edges in dependence of the selected values for $\ell$. The investigation explores all possible values of   $a\in [0,1]$  and $\theta\in[0,\pi/2]$.
The results provide  the constrained celestial coordinate $(\beta,\alpha)$   for  different $\theta$ and  for different spins.  In this way we  relate the photons, within the constraints considered in these analysis,   to the observation coordinates  $(\alpha,\beta)$ (that is the location on the shadow profile), the observation angle  $\theta$
and the  \textbf{BH} spin, distinguishing  points on the \textbf{BHs} shadow profiles correspondent to the fixed constraints. The collections of all the points, for fixed constraints, for all values of the observational angles define curves in the plane $(\alpha,\beta)$ we study in our analysis.

The quantity $\ell_H^\pm(a)$, constraining photons in the \textbf{BH} shadow profile,   are also the basis for  the definition of  replicas\footnote{The quantity $\ell_H^+(a)$ turns out to be significant for the study of \textbf{BH} physics, in particular, to investigate its interaction with the  environment and its  consequent evolution.
 It is a well known fact that limits of \textbf{BH} energy extraction
are imposed   by the \textbf{BH} horizon. Indeed,
massive  particles  or photons with  momentum $p^\mu$ that cross the outer horizon $r_+$
 of a Kerr \textbf{BH}  should satisfy the inequality
$-p_\mu  (\mathcal{L}_H^+)^\mu\geq 0$. This implies that
$\mathcal{E}-\omega_H^+ \mathcal{L}\geq 0$,
for $\mathcal{E}\equiv -p_\mu \xi^\mu_{(t)}$ (energy),
where $\mathcal{L} = p_\mu \xi^\mu_{(\phi)}$ is the  $\phi$ component of the particle (photon) angular momentum.
Thus,
$
\ell\equiv \mathcal{L}/\mathcal{E}\leq \ell_H^+\equiv{1}/{\omega_H^+}$.
If the energy $\mathcal{E}$ is negative,  $\mathcal{L}$ is negative and  the  \textbf{BH} spin is reduced,  (\textbf{BH} spin--down). The process is regulated by
${\delta J}/{\delta M}\leq \ell_H$
(where $ J$ is the Kerr \textbf{BH} spin and $M$ its mass, $\delta$ is for the variation)
regulating also the super-irradiance  as  the  analogue Penrose process for  radiation scattering  by a
Kerr \textbf{BH}.
For a wave--mode of angular-frequency $\omega$,
$\omega$ is amplified if  $\omega\in]0,{z}/{\ell_H^+}[$,
where  $z\in Z-\{0\}$ is  the wave angular momentum number.
}.
  The Kerr  \textbf{BH} events horizons are Killing horizons of $\mathcal{L}_H^\pm$,
  where
$\mathcal{L}_H^\pm\equiv \xi_{(t)} +\omega_H^{\pm} \xi_{(\phi)}$,  $\omega_H^{\pm}$ is the angular velocity {(frequency)} of the outer and inner  horizons, respectively, $(\xi_{(t)},\xi_{(\phi)})$ are the Killing vectors of the Kerr metric, associated with time translations and rotations along the angle $\phi$.
There is  $\ell_H^\pm(a)=1/\omega_H^\pm$.
  The horizons
degenerate at $r=M$ for extreme Kerr \textbf{BH}s, where the angular  frequency is $\omega_H^\pm=1/2$ for a static observer at infinity.
Replicas are  photon orbits whose constant  parameter $\ell\equiv 1/\omega$ is,  in magnitude, $\ell_H^+$ or $ \ell_H^-$.
The set of all  photon  (circular) orbits with equal   fixed    $\omega$,  constitutes an object called metric Killing  bundles (or simply  metric bundles), with characteristic frequency $\omega$, which  is always a \textbf{BH} horizon frequency
 \cite{bundle-EPJC-complete,remnants,Pugliese:2022vni,Pugliese:2022xry,Pugliese:2021aeb,Pugliese:2021hfl,Pugliese:2020azr}. Only some replicas are solutions of system $\mathfrak{(R)}$ and will appear consequently  on the luminous texture of the shadow boundary.

Future observational advancements  and new data analysis techniques will  allow  more clear observation and description of the \textbf{BH} environment, discerning   more refined structure of the images. There is indeed  a continuous improvement (and updating) of the     images.  For example,
 combining  data from   radio telescopes
Global Millimetre \textbf{VLBI} Array, \textbf{ALMA}, and the Greenland Telescope,   the  ring--like accretion structure in \textbf{M87} has been shown connecting the  \textbf{BH}  plasma jet  to the  \textbf{BH} and the accretion matter \cite{Lu23}.
On the other hand a new   analysis of \textbf{EHT} data, using   a    PRIMO algorithm, sharpened the  earlier  view of the glowing gas around the  \textbf{M87} \textbf{BH} \cite{Medeiros}.
In \cite{ccdpp} again  the  \textbf{M87*} images from \textbf{EHT} collaboration  has been  re-analyzed, using  a series of kinetic plasma simulations,
 predicting   the  images during the   outbursts  characterizing \textbf{M87*}.
  New analysis  also provides evidence of a rotating jet ejected from the \textbf{BH} region
--see also \cite{2022MNRAS.517.2462L} and   \cite{Medeiros,PalumboWong,Avery, Avery2,W22SafAstar,2022ApJ...940..182T,Tamburini,2022MNRAS.517.2462L}.
 (Then, in  \cite{Avery},   revisiting  the past observations, it has been  claimed to have spotted  the sharp light ring   around  \textbf{M87*}--see also \cite{Medeiros,PalumboWong, Avery2,Johnson,2022MNRAS.517.2462L,L2, 2023ApJ...944...55P}).

Therefore we  expect that it will be possible to observe  more refined and clearer details from the astronomical observations in near future (see  \cite{Lu23}) and the features  proposed in this work, related to   horizon \textbf{BH} replicas, could  be used as a guiding idea for future observational enhancements where   the regions highlighted here will be distinctly recognizable.

\medskip

This article is structured as follows:

 In Sec.\il(\ref{Sec:quaconsta}), we introduce the spacetime metric and the equations of motion. Constants of motion and geodesic equations are presented in Sec.\il(\ref{Sec:carter-equa}).
The main aspects of black hole shadows are discussed in Sec.\il(\ref{Sec:shadows}).
The constants of motion $q$ and $\ell$ are investigated in Sec.\il(\ref{Sec:q-lell-quantities}).
The celestial coordinates $(\alpha,\beta)$, which define the  shadow boundary, are introduced in Sec.\il(\ref{Sec:alpha-beta}).
In Sec.\il(\ref{Sec:Metric-bundles}) we  discuss the constraint adopted  in our analysis.
 In Sec.\il(\ref{Sec:lblu}) starts the analysis of the \textbf{BH} shadow equations within  condition $\ell=\mp \ell_{H}^\pm$.

We proceed solving equations $(\mathfrak{R})$ within the constraints  on the impact parameter $\ell=\mp \ell_{H}^\pm$.
Hence, in Sec.\il(\ref{Sec:first-lun}), we solve  $\mathfrak{(R)}$   with $\ell=\pm\ell_H^\pm$ for the  radius $r$ and the Carter constant $q$, and in Sec.\il(\ref{Sec:shadows-replicas}) we discuss the celestial coordinates $(\alpha,\beta)$  for the  replicas on the shadow profile.
(Further details are also discussed in Appendix\il(\ref{Sec:further-notes})).
Concluding remarks are given in Sec.\il(\ref{Sec:conclu-Rem}).
Two Appendix sections follow.
More notes on the concept and formalism of metric bundles are in Sec.\il(\ref{Sec:MB-Appendix}):  in Sec.\il(\ref{Sec:MB-Appendix1}) the concept of metric bundles and replicas is  deepen.
Relations between definition of  photon circular  orbits  and  replicas are discussed in Sec.\il(\ref{Sec:photon}).
 Sec.\il(\ref{Sec:photon-shere}) clarifies  the relation between  concepts of  photons spheres   and replicas. In Sec.\il(\ref{Sec:photonu}) we relate circular photons orbits and replicas.
Sec.\il(\ref{Sec:further-notes}) contain details on shadow analysis of Sec.\il(\ref{Sec:first-lun}).
\section{The spacetime metric}\label{Sec:quaconsta}
The   Kerr  spacetime metric  reads
%
\bea \label{alai}&& ds^2=-\left(1-\frac{2Mr}{\Sigma}\right)dt^2+\frac{\Sigma}{\Delta}dr^2+\Sigma
d\theta^2+\left[(r^2+a^2)+\frac{2M r a^2}{\Sigma}\sin^2\theta\right]\sin^2\theta
d\phi^2
-\frac{4rMa}{\Sigma} \sin^2\theta  dt d\phi,
\eea
in the Boyer-Lindquist (BL)  coordinates
\( \{t,r,\theta ,\phi \}\)\footnote{We adopt the
geometrical  units $c=1=G$ and  the $(-,+,+,+)$ signature, Latin indices run in $\{0,1,2,3\}$.  The radius $r$ has unit of
mass $[M]$, and the angular momentum  units of $[M]^2$, the velocities  $[u^t]=[u^r]=1$
and $[u^{\phi}]=[u^{\theta}]=[M]^{-1}$ with $[u^{\phi}/u^{t}]=[M]^{-1}$ and
$[u_{\phi}/u_{t}]=[M]$. %
},
where
\bea\label{Eq:delta}
\Delta\equiv a^2+r^2-2 rM\quad\mbox{and}\quad \Sigma\equiv a^2 \cos^2\theta+r^2.
\eea
 The parameter  $a=J/M\geq0$ is the metric spin, where the  total angular momentum is   $J$  and  the  gravitational mass parameter is $M$.
  A Kerr \textbf{BH} is defined  by the condition $a\in[0,M]$  with inner and outer Killing horizons $r_-\leq r_+$ respectively, where  $r_{\pm}\equiv M\pm\sqrt{M^2-a^2}$.
  The extreme Kerr \textbf{BH}  has dimensionless spin $a/M=1$ and the non-rotating   case $a=0$ corresponds to  the   Schwarzschild \textbf{BH} solution.
  The Kerr naked singularities (\textbf{NSs}) have  $a>M$.

For further analyses it is convenient to introduce the angular parameter $ \sigma\equiv \sin^2\theta\in[0,1]$.
The equatorial plane, $\sigma=1$, is a metric symmetry plane  and   the  equatorial  circular geodesics are confined to the equatorial  plane as a consequence of the metric tensor symmetry under reflections with respect to the plane $\theta=\pi/2$.
The outer ergoregion of the spacetime is  $]r_+,r_{\epsilon}^+
]$, where the  outer  stationary  limit $r_{\epsilon}^+$ (outer ergosurface) is
\bea\label{Eq:sigma-erg}
r_{\epsilon}^{+}\equiv M+\sqrt{M^2- a^2(1-\sigma)},
\eea
where
  $r_{\epsilon}^+=2M$  on the equatorial plane $\theta=\pi/2$ ($\sigma=1$),
and  $r_+<r_{\epsilon}^+$ on   $\theta\neq0$.

In the following analyses, we will  use  dimensionless   units with $M=1$ (equivalent to $r\rightarrow r/M$  and $a\rightarrow a/M$).

\subsection{Constants of motion and geodesic equations}\label{Sec:carter-equa}

We consider  the  following  constants of  motion
\bea&&\label{Eq:EmLdef}
\Em=-(g_{t\phi} \dot{\phi}+g_{tt} \dot{t}),\quad \La=g_{\phi\phi} \dot{\phi}+g_{t\phi} \dot{t},\quad  g_{ab}u^a u^b=\kappa \mu^2,
\eea
which are associated with the  rotational  Killing field   $\xi_{(\phi)}\equiv \partial_{\phi}$
     and  the Killing field  $\xi_{(t)}\equiv \partial_{t}$,
representing the stationarity of the  background. Moreover,
   $u^a\equiv\{ \dot{t},\dot{r},\dot{\theta},\dot{\phi}\}$, where
a dot stands for the derivative with respect to the proper time (for  $\mu>0$) or, in the case of lightlike geodesics,  to  a properly defined  affine parameter (for $\mu=0$). Finally,   $\kappa=(\pm,0)$ is a normalization constant with $\kappa=-1$ for  massive  particles.
 The constant $\La$ in Eq.\il(\ref{Eq:EmLdef}) may be interpreted       as the axial component of the angular momentum  of a test    particle following
timelike geodesics and $\Em$  represents the total energy of the test particle, as measured  by  a static observer at infinity.

We  introduce also  the specific  angular momentum \cite{Bardeen1973,Chandra}
 \bea&&\label{Eq:flo-adding}
\ell\equiv\frac{\La}{\Em}=-\frac{g_{\phi\phi}u^\phi  +g_{\phi t} u^t }{g_{tt} u^t +g_{\phi t} u^\phi} =-\frac{g_{t\phi}+g_{\phi\phi} \omega }{g_{tt}+g_{t\phi} \omega},
\eea
where  $\omega \equiv{u^\phi}/{u^{t}}$ is the relativistic angular velocity, related to the parameter  $\ell$ as follows
\bea
 \omega(\ell)= -\frac{g_{t\phi}+ g_{tt} \ell}{g_{\phi\phi}+ g_{t\phi} \ell}.
\eea

%
In the following we consider co-rotation or counter-rotation according to the
condition $a\ell\gtrless 0$ respectively.  The  geodesic equations  in the Kerr  spacetime are fully separable and define four constants of motion. For convenience  we summarize  the  (Carter) equations  of motion as follows \cite{Carter}:
\bea&&\label{Eq:eqCarter-full}
 \dot{t}=\frac{1}{\Sigma}\left[\frac{P \left(a^2+r^2\right)}{\Delta}-{a \left[a \Em (\sin\theta)^2-\La\right]}\right],\quad
\dot{r}=\pm \frac{\sqrt{R}}{\Sigma};\quad \dot{\theta}=\pm \frac{\sqrt{T}}{\Sigma},\quad\dot{\phi}=\frac{1}{\Sigma}\left[\frac{a P}{\Delta}-\left[{a \Em-\frac{\La}{(\sin\theta)^2}}\right]\right],
\eea
where
\bea\label{Eq:def-Carter-eq}
&& P\equiv \Em \left(a^2+r^2\right)-a \La,\quad R\equiv P^2-\Delta \left[(\La-a \Em)^2+\mu^2 r^2+\Qa\right],\quad  T\equiv \Qa-
(\cos\theta)^2 \left[a^2 \left(\mu^2-\Em^2\right)+\left(\frac{\La}{\sin\theta}\right)^2\right],
\eea
and
\bea
&&\label{Eq:eich}
\Qa=(\cos\theta)^2 \left[a^2 \left(\mu^2-\Em^2\right)+\left(\frac{\La}{\sin\theta}\right)^2\right]+(g_{\theta\theta} \dot{\theta})^2
\eea
is the Carter constant.
%


\section{Shadows}\label{Sec:shadows}
In this section, we introduce the concept of \textbf{BH} shadows. A description of the constants of motion $(q,\ell)$ is presented in Sec.\il(\ref{Sec:q-lell-quantities}).
The celestial coordinates $(\alpha,\beta)$ are discussed in Sec.\il(\ref{Sec:alpha-beta}).
In Sec.\il(\ref{Sec:Metric-bundles}) we  discuss the constraints adopted  in our analysis.
\subsection{The quantities $(q,\ell)$}\label{Sec:q-lell-quantities}
According to the null  geodesic
equations (\ref{Eq:eqCarter-full}), the boundary (edge, or luminous edge) of the
\textbf{BH} shadow, as described by \cite{Bardeen1973}, is  determined by the (unstable) photon orbits  defined by\footnote{In this work, we also consider solution $ \partial_r^2R=0$. Note that $R$ does not depend explicitly on $\sigma$.}
\bea\label{Eq:radial-condition}
R=\partial_r R=0,\quad \partial_r^2R>0
\eea
(hereafter refereed as set $(\mathfrak{R})$ of equations),  with
constraints provided by the  set of Eqs.\il(\ref{Eq:eqCarter-full}) (in particular the condition $T\geq 0$).
Following the usual procedure, we introduce   the two impact parameters
\bea\label{Eq:lq-constants}
\ell\equiv \frac{\La}{\Em},\quad q\equiv \frac{\Qa}{\Em^2}.
\eea
 for an observer located  at  infinity, where photons
arrive close to the  equatorial plane.
Using Eqs\il(\ref{Eq:radial-condition}), we can find  the impact parameters $(\ell,q)$,  
  independently of $\sigma$. 
Clearly, the condition
($R=0$, $R'=0$) for  $a=0$ and  $q=0$ corresponds to  $(r=3, \ell=\pm 3\sqrt{3})$, that is, the last lightlike circular orbit on the equatorial plane of the Schwarzschild spacetime, where
$R''>0$.
\subsection{The celestial coordinates $(\alpha,\beta)$}\label{Sec:alpha-beta}

Let us introduce the celestial coordinates
$\alpha$ and $\beta$ as follows \cite{Bardeen1973}:
\bea\label{Eq:alpha-beta-first}
\alpha\equiv\lim\limits_{r_o\to+\infty}\left(-r_o^2 \sin\theta_o \left.\frac{d\phi}{dr}\right|_o\right),\quad \beta\equiv\lim\limits_{r_o\to+\infty}\left(r_o^2 \left.\frac{d\theta}{dr}\right|_o\right).
\eea
 We assume that the observer is at infinity and $r_o$ is the  distance from the central attractor to the position of the observer ($r$ is the ``emission" point),   $\theta_o$ is  the inclination angle (the angular coordinate of the distant observer  $\theta_0$ is the inclination angle between the line of
sight of the distant observer and the axis of rotation of the central gravitating object). Therefore, $\alpha$ and $\beta$  are the apparent perpendicular distances
of the shadow, as seen from the axis of symmetry, and its projection on the
equatorial plane, respectively.
The parameters
$(\alpha,\beta)$ are also  called the  horizontal and vertical
impact parameters respectively and are, therefore, evaluated with respect to the point of view of an observer located at infinity .
In the following, for simplicity,   the subindex $o$  in $(r_o,\theta_o)$  will be dropped.

Using  Eqs\il(\ref{Eq:eqCarter-full}) for  null geodesics  in Eqs\il(\ref{Eq:alpha-beta-first}), we obtain:
\bea&&\label{Eq:beta-d-defin}
\beta=\pm\sqrt{q-q_c}\quad \mbox{for}\quad
\sigma\neq0\quad\mbox{and}\quad  q\geq q_c,\quad\mbox{where}\quad
 q_{c}\equiv\frac{(\sigma -1) \left(a^2 \sigma -\ell ^2\right)}{\sigma},
 \eea
%
where $\sigma\neq0$ and  in particular
 for $ \sigma=1$, there is $ \beta=\pm\sqrt{q}$.

The quantity   $\beta$ is independent of the rotation orientation (sign of $\ell$ sign) and of the inclination angle orientation (it is even on $\theta$). For $\sigma=0$ (attractor poles), this quantity is not well defined.  On the equatorial plane, it has to be $q\geq0$. The sign of $\beta$ depends on the sign of $(\dot{\theta},\dot{r})$   in Eqs\il(\ref{Eq:alpha-beta-first}).

The coordinate $\alpha $ satisfies the relationship
\bea\label{Eq:alpha-def}
\alpha=-\frac{\ell}{\sin\theta},\quad \mbox{and }\quad\alpha=-\ell\quad \mbox{for}\quad \theta=\frac{\pi}{2}.
\eea
Note that $\alpha $ depends on the sign of $\ell$,  is odd in  $\theta$, and does not depend on the  attractor spin. We will use
definition Eq.\il(\ref{Eq:alpha-def}) in the form
$\alpha=-{\ell}/\sqrt{\sigma}$, which is equivalent to considering  $\theta\in [0,2\pi]$, using the coordinates
$(\alpha,-\alpha)$.  %
Then,  $\alpha=0$ for $\ell=0$.

In the analysis of shadows, we shall consider the function $\beta(\alpha)$.

The quantities $(q,r)$ do not depend on the angle $\sigma$. However,  the  coordinates with respect to the observer at infinity $(\alpha,\beta)$  depend on  $\sigma$.

\subsection{Constraints}\label{Sec:Metric-bundles}
In this section we  introduce the constraints adopted  in our analysis, for the set  $(\mathfrak{R})$ of Eq.\il(\ref{Eq:radial-condition}).
\textbf{BH} shadow profile is in general found, at fixed spin and observational angle, considering photons coming from \emph{all} the points at
infinity, having  \emph{all} the possible values of the impact parameters $\ell$, therefore solving $(\mathfrak{R})$   for \emph{all} values of $\ell$.

We study solutions of the equations $\mathfrak{(R)}$  in dependence on $a\in [0,1]$  and of the
  $\theta$ angle,   within the constraint  $\ell=\pm \ell_H^\pm(a)$.  That is  $|\ell|$ is evaluated on the \textbf{BH} outer and inner horizons with  $\ell_H^\pm(a)=1/\omega_H^{\pm}(a)$,   where  $\omega_H^{\pm}$ is the angular velocity {(frequency)} of the horizons,  representing   the \textbf{BH} rigid rotation.
  Functions $\ell_H^\pm(a)$ are shown in Figs\il(\ref{Fig:Plotmescoll})--\emph{left panel} as function of the \textbf{BH} spin $a$.  
Therefore, counter-rotating and  co-rotating photons with $\ell=\pm \ell_H^\pm(a)$ are considered  for all possible values of   $a\in [0,1]$  and $\sigma\in [0,1]$ ($\theta\in[0,\pi/2]$).
Hence, these special null circular  orbits,   characterized by  specific angular momentum
$\ell=\pm  \ell_H^\pm$, i.e.  equal in magnitude to the angular momentum of the outer or  inner \textbf{BH} horizons,  have been called  counter-rotating or co-rotating,  outer or inner,  horizons replicas (replicas),\cite{bundle-EPJC-complete,remnants,Pugliese:2022vni,Pugliese:2022xry,Pugliese:2021aeb,Pugliese:2021hfl,Pugliese:2020azr}.
The results for ($\mathfrak{R}$) with  $\pm\ell_H^\pm(a)$ provide  the constrained celestial coordinate $(\beta,\alpha)$,  for  different $\theta$ and  for different spins $a$, distinguishing  parts, points or regions, of the luminous edges (shadows profiles) associated to the  fixed constraints, i.e.
the analysis of solutions  ($\mathfrak{R}$) for $\pm\ell_H^\pm(a)$  individuates   particular replicas that appear  on the luminous profile of the shadow contour. Solutions with the  constraints $\ell=\pm  \ell_H^\pm$ are points on the shadow edge and
  introduce new observable related to   the replicas.
Horizons replicas  can be applied to  extract information about the \textbf{BH} and, particularly, the \textbf{BH} horizons,  by  measuring  properties of the  \textbf{BH} horizons in regions accessible to distant observers \cite{Pugliese:2021aeb,remnants}.  It has been proved in  \cite{bundle-EPJC-complete,remnants,Pugliese:2022vni,Pugliese:2022xry,Pugliese:2021aeb,Pugliese:2021hfl,Pugliese:2020azr} there are  Kerr inner horizons replicas in  regions very close to the \textbf{BH} rotational axis. Therefore, one of the intriguing   applications of  replicas in \textbf{BH} physics is the possibility to explore the regions close to the \textbf{BH} rotational axis in order to obtain information about the horizons\footnote{
From the phenomenological view point, an  observer can  detect  the presence of a replica at the point $p$ of the \textbf{BH} spacetime with spin $a$, by measuring the \textbf{BH} horizon frequency   $\omega_H^+(a)$ or $\omega_H^-(a)$  (the outer or inner \textbf{BH} horizon frequency)  at the point $p$ \cite{Pugliese:2021aeb}.
It has been  pointed out that \textbf{MBs}  appear to be related to the concept of
spacetime \emph{pre-horizon regimes},
indicating the  existence of detectable  mechanical effects allowing circular orbit  observers to
recognize the close presence of an event horizon.
Pre-horizon regimes  were introduced in  \cite{de-Felice1-frirdtforstati,de-FeliceKerr,
de-Felice-anceKerr}.
An implication of this property is that an object like a gyroscope could observe
 a connected phenomenon and  interpret  it as
 a \emph{memory} of the static (Schwarzschild-like) case
in the Kerr metric--\cite{de-Felice-first-Kerr}. Moreover, these structures could play an important role
for the description of  the  black hole formation and for testing the possible existence of naked  singularities and  could be of  relevance during the gravitational
collapse \cite{de-Felice3,de-Felice-mass,de-Felice4-overspinning,Chakraborty:2016mhx}.
 Interpretations of similar related concepts  have been presented also  in \cite{Tanatarov:2016mcs,Mukherjee:2018cbu,Zaslavskii:2018kix}.
  }.   While replicas   reveal their significance in particular  in the region proximal  to $\theta\approx0$ (the \textbf{BH} poles) where they are more numerous--\cite{bundle-EPJC-complete,remnants,Pugliese:2022vni,Pugliese:2022xry,Pugliese:2021aeb,Pugliese:2021hfl,Pugliese:2020azr},
in general the existence of  replicas  is dependent on the \textbf{BH} spin $a$, the radius $r$ and the angle $\theta$.
Replicas, like the critical curve, are a characteristic of spacetime only,  depending  exclusively on the  dimensionless \textbf{BH} spin, therefore they can be considered to describe the characteristics of the central \textbf{BH} from the luminous profile.
Since the  replicas we consider here are \emph{also} solutions of the  equations $\mathfrak{(R)}$, this implies that,   at fixed  spin $a$ and observational angle $\sigma$, they appear as points or regions on the  luminous boundary of the \textbf{BH} shadow, constituting the   luminous texture of the shadow  boundary  profile\footnote{Bundle formalism can easily be related to the solutions of ($\mathfrak{R}$) in the plane $(\alpha,\beta)
$ in the fact  the celestial coordinate  $\alpha$  can be directly related to some bundle characteristics (the bundle origin and the bundle tangent spin)--see discussion in Sec.\il(\ref{Sec:MB-Appendix1}).}.
These orbits having    same $\ell$ in absolute value as the \textbf{BH} horizons allow to map the critical curve (the shadow edge), by  connecting a specific value of the photon impact parameter to certain regions of the shadow contour curve,  the  radius and Carter constant $(r,q)$ and  observational angle $\theta$.

\medskip

\textbf{Notes on replicas and Killing metric bundles}

While the quantity  $\ell_H^+$  is an important \textbf{BH} feature   limiting for example  \textbf{BH} energy extraction, we introduce here  the constraints  $\ell=\pm\ell_H^\pm(a)$ for the system $(\mathfrak{R})$,   as related to the concept of  Killing metric bundles (\textbf{MBs})\cite{bundle-EPJC-complete,remnants,Pugliese:2022vni,Pugliese:2022xry,Pugliese:2021aeb,Pugliese:2021hfl,Pugliese:2020azr}.

In general $\MBs$ are defined from    the more general Killing field
$\mathcal{L}_\omega\equiv \xi_{(t)} +\omega\xi_{(\phi)}$,  where $(\xi_{(t)} \equiv \partial_t,\xi_{(\phi)}\equiv \partial_\phi)$ are metric Killing field, defining the quantity  $\mathbf{\mathcal{L_N}}\equiv\mathcal{L}_\omega\cdot\mathbf{\mathcal{L}}_\omega$, null  for photon-like
particles with particular  rotational frequencies $\omega=\omega_{\pm}$ ($\omega_H^\pm$ are  frequencies $\omega_{\pm}$ evaluated on the \textbf{BH} horizons $r_\pm$, this point  is discussed more extensively in Sec.\il(\ref{Sec:MB-Appendix1})). Therefore $\MBs$ are  solutions of  $\mathcal{L}_{\mathcal{N}}=0$, under certain conditions as $\omega=$constant or $\ell=$constant,  where
$\mathcal{L}_H^\pm=\xi_{(t)} +\omega_H^{\pm} \xi_{(\phi)}$, and
the \emph{angular momentum}  $\ell_H^\pm\equiv 1/\omega_H^\pm$.  In this analysis we look for those particular solutions that are also solutions of $\mathfrak{(R)}$.

In general, the set of all  solutions $\mathcal{L}_{\mathcal{N}}$ with \emph{equal   fixed} orbital     $\omega$ (bundle characteristic frequency),  constitutes a metric Killing  bundle (or simply  metric bundle). Constant $\omega$  turns to be  always a \textbf{BH} horizon frequency, hence bundle orbits are, in general, all replicas of \emph{a} \textbf{BH} horizon.
 (Example of replicas on the equatorial plane of Kerr \textbf{BH} with spin $a\in[0,1]$ are discussed in Sec.\il(\ref{Sec:MB-Appendix1})--(see Eq.\il(\ref{Eq:Eqmescol}) and Fig.\il(\ref{Fig:Plotmescol}))). However
we are interested here in the case of replicas in a \emph{fixed}  Kerr \textbf{BH} background.
More notes on the concept and formalism of $\MBs$ are in Sec.\il(\ref{Sec:MB-Appendix}). In particular, in Sec.\il(\ref{Sec:MB-Appendix1}) the concept of metric bundles and replicas is  deepen, considering in particular the relation with photon circular geodesic obits.
Relations between definition of  photon circular  orbits  and  replicas are discussed in Sec.\il(\ref{Sec:photon}).
 Sec.\il(\ref{Sec:photon-shere}) clarifies  the relation between  concepts of  photons spheres   and replicas. In Sec.\il(\ref{Sec:photonu}) we relate circular photons orbits and replicas.

%
 \section{The condition $\ell=\epsilon \ell_{H}^\pm$}\label{Sec:lblu}
We proceed solving equations $(\mathfrak{R})$ within the constraints  on the impact parameter $\ell=\epsilon\ell_H^\pm$, where $\epsilon=\{+1,-1\}$,  examining  the associated  regions on the shadow profiles, in dependence on the \textbf{BH} spin and the observational angle.
Results  provide, for each constraint, the coordinate $(\beta,\alpha)$, in dependence of ranges of  \textbf{BH} spin $a$, angle  $\sigma$, and  Carter constant $q$.
 Constrained photons  can be traced on specific regions of the luminous edge  and  vary depending on the angle of observation and spin. We obtain a  relation $\beta(\alpha)$ i.e.   we can relate  points or regions  of  the luminous profile (at fixed $\theta$ and $a$) to the fixed constraints.
Hence, in Sec.\il(\ref{Sec:first-lun}), we solve  $\mathfrak{(R)}$   with $\ell=\epsilon\ell_H^\pm$ for the variables $(r,q)$, considering the constraints imposed by the Carter equations  and in Sec.\il(\ref{Sec:shadows-replicas}) we discuss the celestial coordinates $(\alpha,\beta)$  for the  replicas on the shadow profile.
Further details are discussed in Appendix\il(\ref{Sec:further-notes}).
\subsection{Quantities $(q,r)$ for $\ell=\epsilon \ell_{H}^\pm$}\label{Sec:first-lun}
In this Section we constrain the quantities $(q,r)$,  solutions of  $\mathfrak{(R)}$ with $\ell=\epsilon\ell_H^\pm$, for different $(a,\sigma)$.

That is we consider the Carter equations for the \textbf{BH} shadow profiles with the constraints  $\ell=\epsilon\ell_H^\pm$, discussing the results in terms of the quantities $(r,q)$.  By  constraining   the spin $a$,  regions $r$, constant $q$ and  observational angle $\theta$,  results provide, for each constraint  on the parameter $\ell$,  the  replicas which are also  solutions of the shadow equations $(\mathfrak{R})$, hence  forming  the luminous edge solution of $(\mathfrak{R})$.  In  Sec.\il(\ref{Sec:shadows-replicas}) we discuss the results in terms of the coordinate celestial coordinates $(\alpha,\beta)$.

Let us note there are no solutions of  $\mathfrak{(R)}$  $(T>0)$ with  $q<0$ and $R=0$ (see Eq.\il(\ref{Eq:beta-d-defin}) and (\ref{Eq:def-Carter-eq}));
therefore, there are no solutions for the shadow problem with
$q<0$  so that this condition has to be eliminated from our analysis. 

\medskip

\begin{figure}[h]
\centering
\includegraphics[width=8cm]{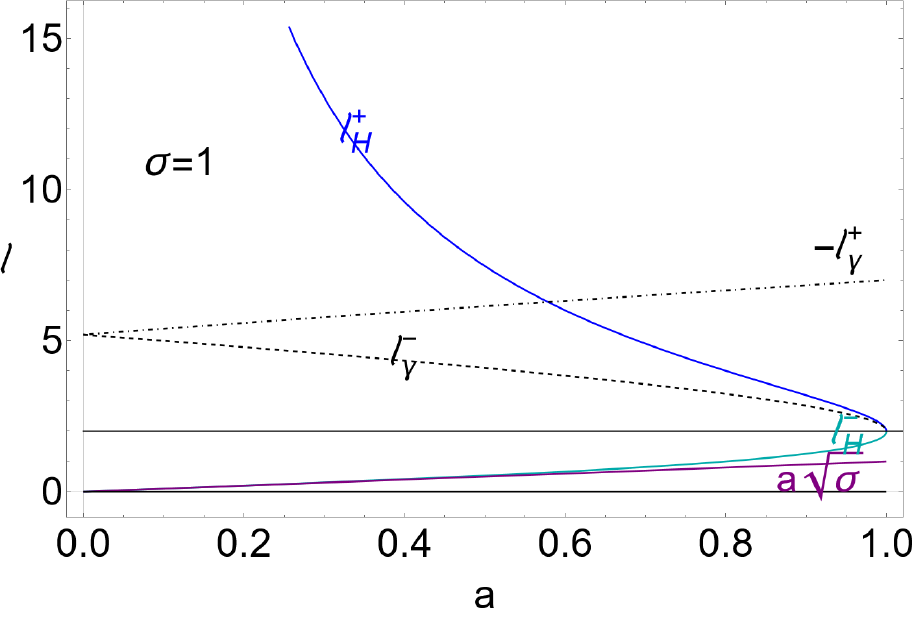}
\includegraphics[width=9cm]{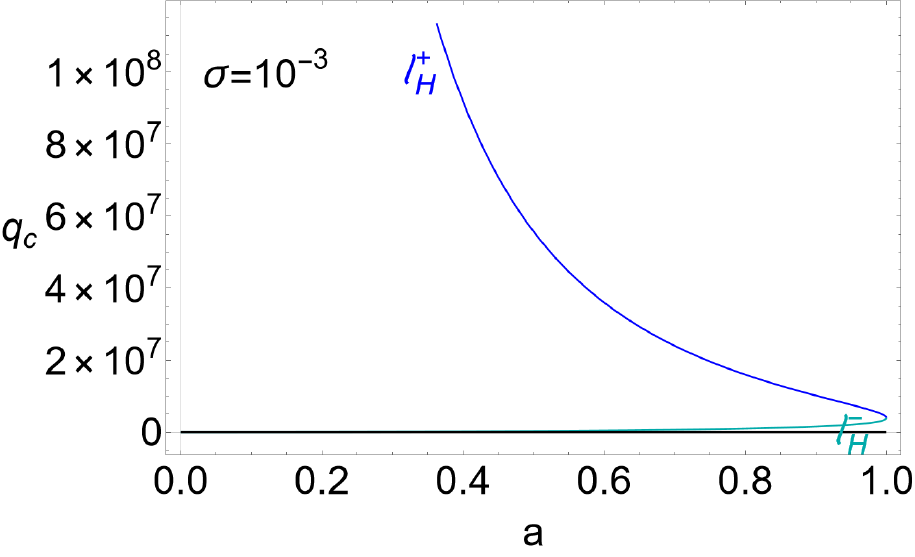}
\caption{Left panel: Specific angular momentum $\ell_H^\pm(a)$  of the outer and inner  Kerr \textbf{BH} horizons, respectively, as functions of the \textbf{BH} dimensionless  spin $a/M$ and the limiting parameter $\ell=a\sqrt{\sigma}$ (purple line) on the equatorial plane $\sigma\equiv \sin^2\theta=1$.  Radii $r_\gamma^\pm$ are the photon counter-rotating and co-rotating  circular orbits respectively  on the equatorial plane (and  boundary of the Kerr \textbf{BH} photon sphere--see Sec.\il(\ref{Sec:photon-shere}). Right panel: Function $q_c$ of Eq.\il(\ref{Eq:beta-d-defin})  ($q$ is the Carter constant)  for $\sigma=10^{-3}$ and evaluated for $\ell=\ell_H^\pm(a)$ as function of the \textbf{BH} dimensionless spin. All quantities are dimensionless.}.\label{Fig:Plotmescoll}
\end{figure}

Considering  Eq.\il(\ref{Eq:beta-d-defin}) for   $q>0$,  the condition $T\geq0$ implies $q\geq q_c$, that is,
$\sigma\in]0,1[$  with $\ell^2<a^2\sigma$ ($q_c<0$),  and
 $\ell^2>a^2\sigma$  ($q_c>0$) with $q>q_c$. Function $q_c$, evaluated on $\ell_H^\pm$ is represented  in  Figs\il(\ref{Fig:Plotmescoll})--right panel, as functions of $a$

 Also, on the equatorial plane
 $\sigma=1$ (or for $\ell^2=a^2\sigma$, i.e., $\alpha^2=a^2$) we obtain   $q_c=0$ and  $q>0$.

For $q=0$, from Eq.\il(\ref{Eq:beta-d-defin}) we obtain that $\sigma=1$ ($q_c=0$ and, therefore, also the condition $\ell^2=a^2 \sigma$), or  $\sigma\in ]0,1[$ with $\ell^2<a^2 \sigma$ $(q_c<0)$.

\begin{figure}[h]
\centering
\includegraphics[width=8cm]{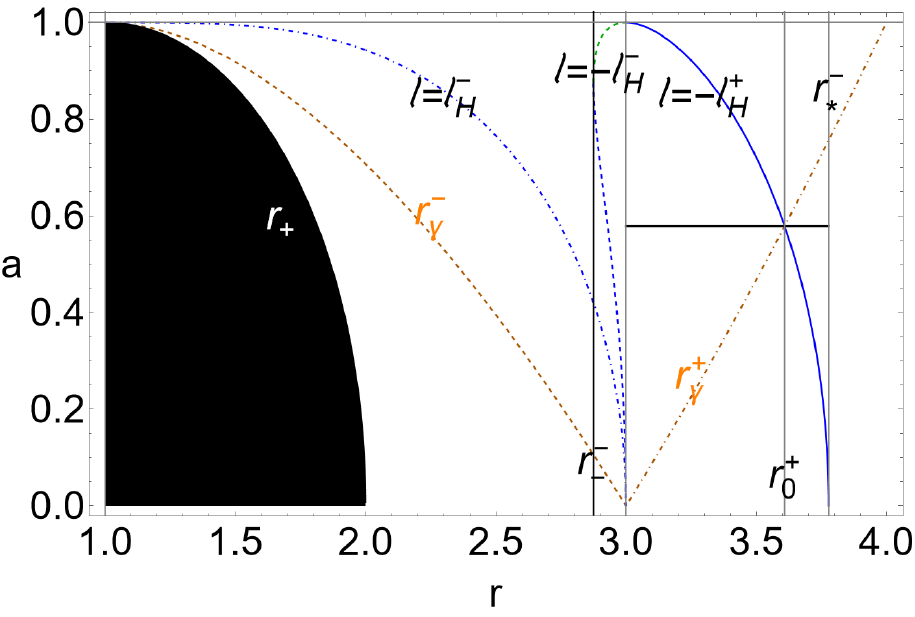}
\includegraphics[width=8cm]{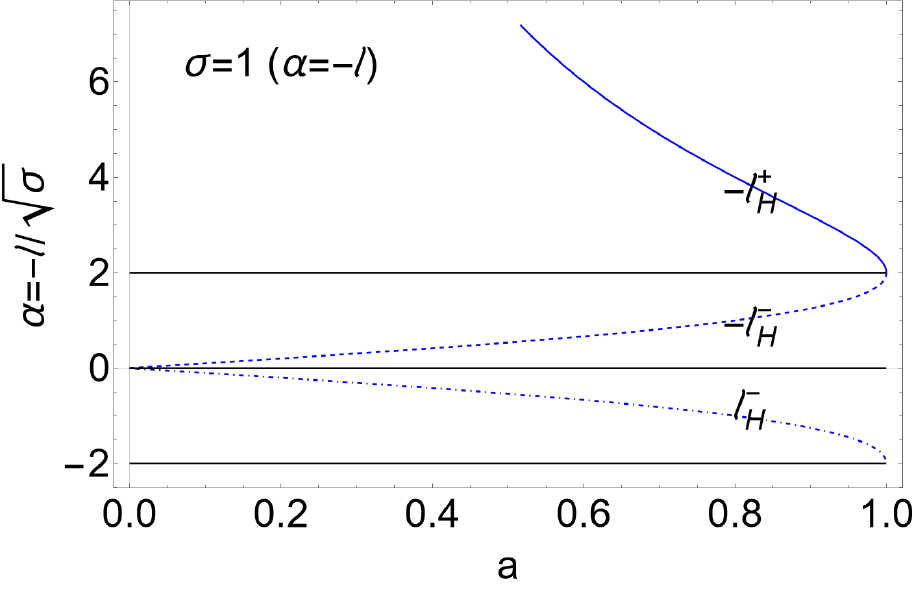}
\caption{Left panel: Curves show the relations $(a,r)$ in the solutions of the equations $(\mathfrak{R})$ of Eq.\il(\ref{Eq:radial-condition})  for  $\ell=\pm\ell_H^\pm$.  $\ell_H^\pm$ are the specific angular momenta of the outer and inner Kerr \textbf{BH} horizons, respectively. The \textbf{BH} region, shown in black, is bounded by the function $a: r=r_+$, where $r_+$ is the \textbf{BH} horizon.   Spin function $a(r)$ are  function of the  dimensionless radius $r$  for $\ell=-\ell_H^+$ (blue solid curve),  $\ell=\ell_H^-$ (blue dotted-dashed curve) and  $\ell=-\ell_H^-$ (blue dashed curve).  Radii  $\{r_-^-,3,r_0^+,r_*^-\}$ bounding $r$ are also represented. Further discussion on the $(r,a)$ in these cases is also in Sec.\il(\ref{Sec:further-notes}).
Radii $r_\gamma^\pm$ (orange dotted and dotted-dashed lines) are the photon counter-rotating and co-rotating  circular orbits respectively  on the equatorial plane (and  boundary of the Kerr \textbf{BH} photon sphere see Sec.\il(\ref{Sec:photon-shere}).).  Right panel: celestial coordinate $\alpha\equiv-\ell/\sqrt{\sigma}$  of Eq.\il(\ref{Eq:alpha-def}) as  function of the \textbf{BH} spin  for $\sigma=1$ (equatorial plane) and $\ell\in\{\pm\ell_H^-,-\ell_H^+\}$. (All quantities are dimensionless.).}\label{Fig:Plotqmminatint}
\end{figure}

From Figs\il(\ref{Fig:Plotmescoll})--left panel it is clear that, for any $\sigma\in]0,1]$,
 $\ell_H^-\in]a\sqrt{\sigma}, \ell_H^+]$.
 (See also Figs\il(\ref{Fig:Plotqmminatint})--right panel, where celestial coordinate $\alpha\equiv-\ell/\sqrt{\sigma}$  of Eq.\il(\ref{Eq:alpha-def})  is shown as  function of the \textbf{BH} spin  on the equatorial plane and $\ell\in\{\pm\ell_H^-,-\ell_H^+\}$.).

 \medskip

Below we discuss in details the results at fixed impact parameter,   considering  solutions of  $\mathfrak{(R)}$  for the cases $\ell=+\ell_H^\pm$ and $\ell=-\ell_H^\pm$ in terms of the constant $q$ and the radius $r$.

\begin{figure}[h]
\centering
\includegraphics[width=8cm]{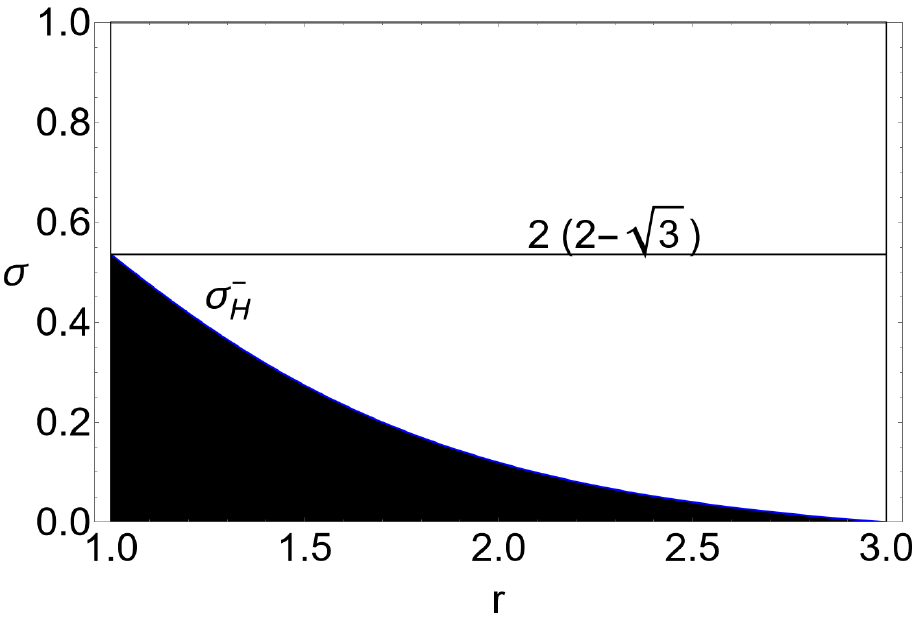}
\includegraphics[width=8cm]{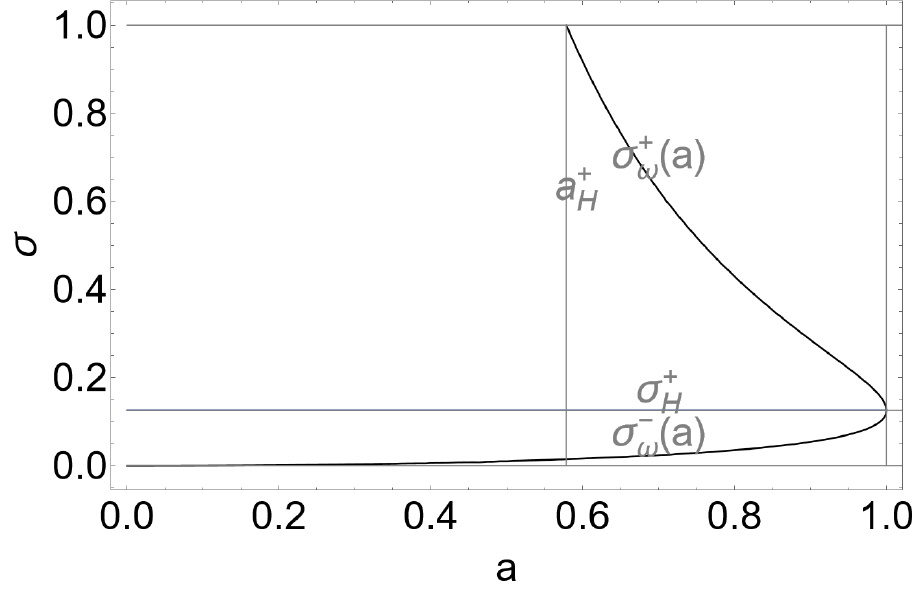}
\caption{Relations $(\sigma,r)$ (left panel) and $(\sigma,a)$ of  the solutions of the equations $(\mathfrak{R})$ of Eq.\il(\ref{Eq:radial-condition})  for  $\ell=\pm\ell_H^\pm$. It follows the discussion of Sec.\il(\ref{Sec:first-lun}).  Quantities $\ell_H^\pm(a)$ are the specific angular momenta of the outer and inner Kerr \textbf{BH} horizons, respectively, $a$ is the dimensionless \textbf{BH} spin,  and  $\sigma\equiv \sin^2\theta\in[0,1]$.   Left panel: Plane $\sigma_H^-$ defined in Eq.\il(\ref{Eq:sigmaHm-def}) as a function of $r$. Solutions   of Eq.\il(\ref{Eq:radial-condition})  for $\ell=\ell_H^-$ exist  for $\sigma\in[\sigma_H^-,1]$. The limiting value $\sigma\approx 0.58$ is also shown. Right panel: Planes $\sigma_\omega^\pm$, solutions of the equation
Eq.\il(\ref{Eq:sigma-omega-eq}), as functions of the \textbf{BH} dimensionless spin. Limiting spin $a_H^+$ is also shown.  All quantities are dimensionless.}\label{Fig:Plotqmminatintsigma}
\end{figure}

\begin{figure}[h]
\centering
\includegraphics[width=8cm]{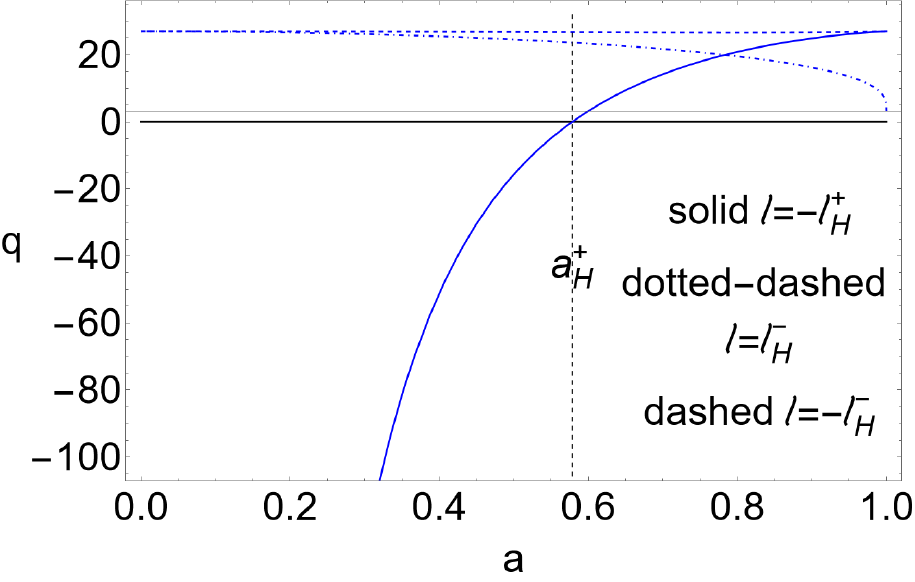}
\includegraphics[width=8cm]{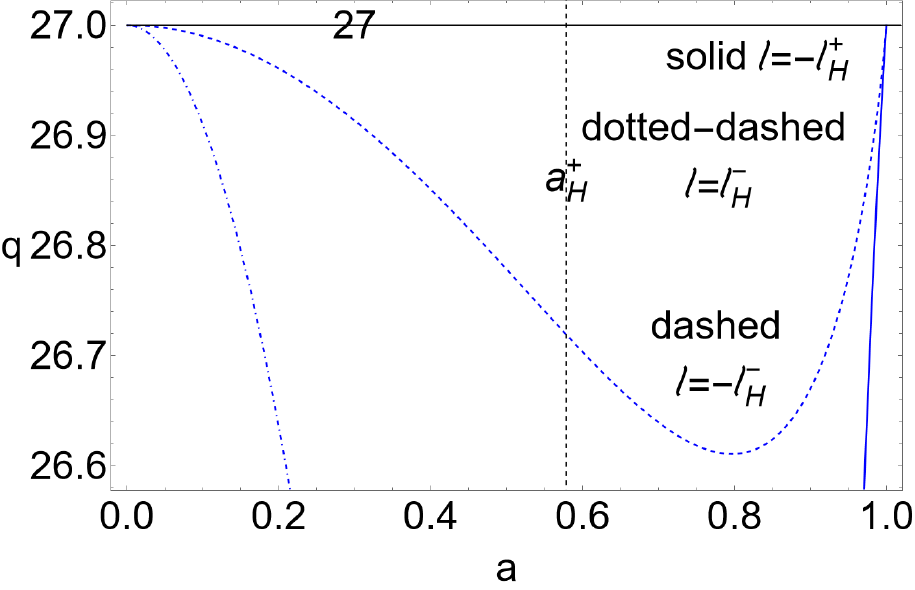}
\includegraphics[width=8cm]{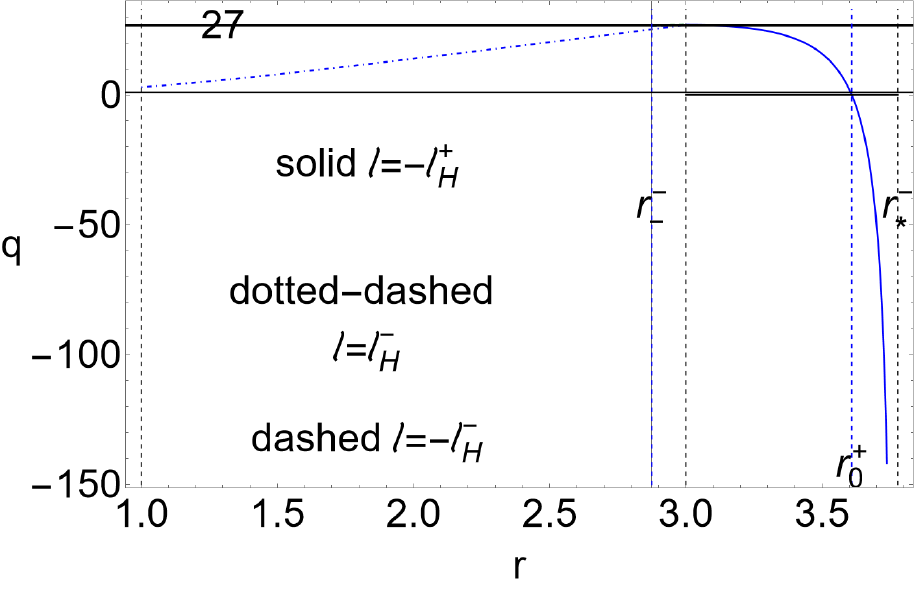}
\includegraphics[width=8cm]{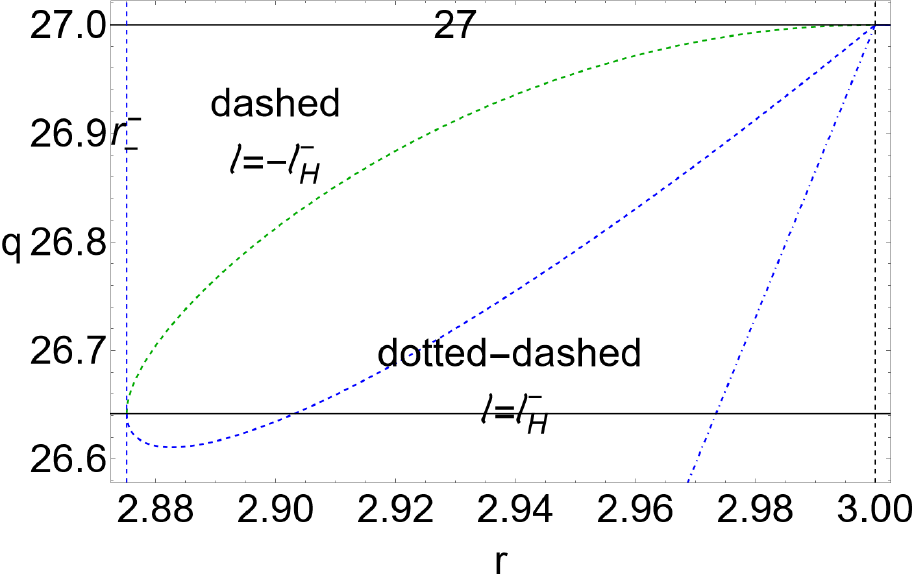}
\caption{Quantity $q$ (Carter constant of Eq.\il(\ref{Eq:lq-constants})) solutions of the set $(\mathfrak{R})$  of Eq.\il(\ref{Eq:radial-condition})  for $\ell=\pm\ell_H^\pm$  as a function of the \textbf{BH} dimensionless spin (upper panels) and  as a function of the  dimensionless radius $r$ (bottom panels)  for $\ell=-\ell_H^+$ (solid curve),  $\ell=\ell_H^-$ (dotted-dashed curve) and  $\ell=-\ell_H^-$ (dashed curve).   Vertical  dashed lines are limiting spins and radii.  The quantities $\ell_H^\pm$ are the outer and inner horizons angular momentum, respectively.   The  right panels are a close--up view of the left panels. The zeros $q=0$  and  upper limit $q=27$ are also shown.  All the quantities are dimensionless. }.\label{Fig:Plotqmminatutt}
\end{figure}

 In general   $(a,r)$ are related  in the solutions of $(\mathfrak{R})$   as shown in
Figs\il(\ref{Fig:Plotqmminatint})--left panel.
Relations $(\sigma,r)$ and $(\sigma,a)$ of  $(\mathfrak{R})$   are in
Figs\il(\ref{Fig:Plotqmminatintsigma}).
Quantity $q$, solution of the set $(\mathfrak{R})$    as a function of the \textbf{BH} dimensionless spin  and  of the  dimensionless radius $r$  are in
Figs\il(\ref{Fig:Plotqmminatutt}).

In details, no solutions  have been found for the system   $\mathfrak{(R)}$   with the constraint  $\ell=\ell_H^+(a)$ (co-rotating outer horizon replicas).
For $\ell=-\ell_H^+$ (counter-rotating outer horizon replicas) we consider first the   solutions of  $\mathfrak{(R)}$  for   $(r,q)$.
A solution is for
%
$q\equiv q_H^+(r,a)$,
where
 \bea&&\label{Eq:qll}
 q_H^\pm(r,a)\equiv \frac{r \left[S_\pm-8r_\pm (r-2)\right]}{a^2\Delta}, \quad \mbox{where}\quad S_\pm\equiv a^2\left\{a^2 (r+2)\pm \left[r(r^2+4)+8 \sqrt{1-a^2}\right]\right\},
 \eea
 %
%
%
for  a radius $r_H^+(a)\in ]3,r_*^-]$ where  $r_*^-\equiv 
3.7784$,
 and $a\in[0,1]$.
 (More in general the relations among $(a,r,q)$ are shown for $\ell=-\ell_H^+$ in Figs\il(\ref{Fig:Plotqmminatutt})--solid blue curves.).

 In Figs.\il(\ref{Fig:Plotqmminatint})--left panel,  relation $(a,r)$ in the solutions of the equations $(\mathfrak{R})$ for  $\ell=-\ell_H^+$ is represented as spin  $a(r): r=r_H^+(a)$, function of the  dimensionless radius $r$  (solid blue curve). Radius  $r_*^-$, bounding $r$ for the extreme Kerr \textbf{BH}, is  also represented.  Radius $r_H^+$ decreases with the \textbf{BH}  spin. {(Further information is in Sec.\il(\ref{Sec:further-notes}))}.

Therefore, considering  null geodetics  with  $\ell=-\ell_H^+$, and   assuming the conditions
 $(R=0, R'=0, R''\geq 0)$ with  $T\geq0$, we obtain   solutions  of $(\mathfrak{R})$  for  the extreme Kerr \textbf{BH} $(a=1, r=3, \sigma \in [\sigma_H^+,1])$, where the limiting observational angle is   $\sigma_H^+\equiv 2 \left(8-3 \sqrt{7}\right)= 0.1255$.

 For \textbf{BHs} with spin  $
a\in ]a_H^+,1]$, where $a_H^+\equiv 0.5785$, there is     $\sigma\in [\sigma_\omega^+(a),1]$  and
$r\in ]3 , r_0^+[$. Finally, for $
a=a_H^+$  there are solutions only on the equatorial plane  $\sigma=1$--see Fig.\il(\ref{Fig:Plotqmminatintsigma})--right panel.
The upper limit on the radius is $ r_0^+\equiv 3.609$ (see Fig.\il(\ref{Fig:Plotqmminatint})--left panel) and
limiting angle
$\sigma_\omega^+$ is shown in Figs\il(\ref{Fig:Plotqmminatintsigma})--right panel  as function of the \textbf{BH} dimensionless spin  ($\sigma_\omega^\pm$ can be  found from
Eq.\il(\ref{Eq:sigma-omega-eq})).

\medskip

 There are solutions of $\mathfrak{(R)}$ for  $\ell=-\ell_H^-$ (counter--rotating inner horizon replica) for all \textbf{BH} spin
$a\in[0,1]$
with  constant
$q=q_H^-$ of Eq.\il(\ref{Eq:qll}),
and  radius $r_H^-(a)\in[r_{-}^-,3]$, where $r_-^-\equiv 2.87513$--see\footnote{Radius $r_H^-(a)$ can be found as a zero of the quantity  $R_H^+$ of Eq.\il(\ref{Eq:RHipiu}).} Figs\il(\ref{Fig:Plotqmminatint})--left panel.
More in general  the quantities $(r,q)$, for different \textbf{BH} spins $a$, are shown for $\ell=-\ell_H^-$ in  Figs\il(\ref{Fig:Plotqmminatutt}). Radius $r$, solution of the equations $(\mathfrak{R})$ for  $\ell=-\ell_H^-$, for all \textbf{BH} spin $a$ is shown  in Figs.\il(\ref{Fig:Plotqmminatint})--left panel,  as  spin function  $a(r): r=r_H^-(a)$, function of the  dimensionless radius $r$ (dashed curve).
Then,
 with  the conditions $(T\geq0,R'=0, R=0)$,
 there is a solution of $(\mathfrak{R})$  in particular  for the Schwarzschild \textbf{BH} with   $(a=0, \sigma\in ]0,1], r=3)$,  for the Kerr \textbf{BH} in
$(a\in ]0,1[, \sigma \in [\sigma_\omega^-,1])$,
and  for the extreme Kerr \textbf{BH} with $(a=1, r=3, \sigma\in [\sigma_H^+,1])$,
where the limiting angles $(\sigma_\omega^-(a),\sigma_H^+)$ are  shown in  Figs\il(\ref{Fig:Plotqmminatintsigma})--\emph{right panel} (and can be found as  solutions of Eq.\il(\ref{Eq:sigma-omega-eq})).

\medskip

Finally, there are solutions of $(\mathfrak{R})$  with
$\ell=\ell_H^-$ (co--rotating inner horizon replicas) for  all  for \textbf{BH} spin  $a\in [0,1]$
 for %
$q=\bar{q}_H^-$, with
\bea
\bar{q}_H^-\equiv \frac{r \left[\bar{S}-8 r_- (r-2)\right]}{a^2\Delta}\quad\mbox{where}\quad \bar{S}\equiv 
S_+-16a^2,
\eea
 %
and $\bar{r}_H^-(a)\in]1,3]$ (where the radius $\bar{r}_H^-(a)$  is a zero of the  quantity $\bar{R}_H^-$ in Eq.\il(\ref{Eq:RHM})).

Quantities $(a,r,q)$ in the solutions of $(\mathfrak{R})$, for $\ell=\ell_H^-$  are related as in  Figs\il(\ref{Fig:Plotqmminatutt})--(blue curves), while in Figs.\il(\ref{Fig:Plotqmminatint})--left panel,  relation $(a,r)$ in the solutions of the equations $(\mathfrak{R})$ for  $\ell=\ell_H^-$ is represented as spin  $a(r): r=\bar{r}_H^-(a)$, function of the  dimensionless radius $r$.
In particular, for $\ell=\ell_H^-$, the solutions are    $r\in [1,3]$ and
$(r = 3, a =0,\sigma\in ]0,1[)$, and for  $r\in ]1,3[$
we obtain $a=\bar{a}_H^-$
(Figs\il(\ref{Fig:Plotqmminatint})--left panel)
with  the observational angle bounded in the range $\sigma \in [\sigma_H^-,1]$,
where
\bea\label{Eq:sigmaHm-def}
\sigma_H^-\equiv-\frac{2 \left(r\psi+6\right)}{(r-3) \chi},\quad\mbox{where}\quad \chi\equiv [(r-1) r+2]^2 ,\quad \psi\equiv [r (3 r-1)+7] r-2 \sqrt{2} \sqrt{\chi [r (r+2)+3]}+1
\eea
--see Figs\il(\ref{Fig:Plotqmminatintsigma})--left panel.

\medskip

In Figs\il(\ref{Fig:Plotqmminatutt})--\emph{upper panels},
 $q$  is shown  as a function of the \textbf{BH} dimensionless spin  for $\ell=-\ell_H^+$  and  $\ell=\pm\ell_H^-$.  The $q$ constant is upper bounded by the value $q=27$.  We analyze, particularly, the case of zero Carter constant ($q=0$)  for $\ell=-\ell_H^+$. This occurs on the background geometry  $a=a_H^+$. (The case $q=0$ reproduces the photon circular orbits on the equatorial plane). The constant $q$ is non negative for $\ell=\pm \ell_H^-$, increases   constantly with the spin for $-\ell_H^+$, and decreases with the spin for $\ell_H^-$. The situation for $\ell=-\ell_H^-$ is more complex and it is shown in Figs\il(\ref{Fig:Plotqmminatutt})--\emph{right panel}.

From   Figs\il(\ref{Fig:Plotqmminatutt}) it can be seen the radial coordinate is, in general, bounded in the range $r\in [1,3.8]$, i.e.,
$r\in [1,3]$ for $\ell=\ell_H^-$,  $r\in [r_-^-,3]$ for $\ell=-\ell_H^-$ (note that in this case there are two values of $q$ for a fixed orbit $r$--see also Figs\il(\ref{Fig:Plotqmminatutt})), and $r\in[3,r_*^-]$ for $\ell=-\ell_H^+$, where $q=0$ occurs for $r=r_0^+$.
The function $r(a)$ is then shown in Figs\il(\ref{Fig:Plotqmminatint}) as the implicit function $a=a(r)$.  For a fixed spin $a$, it is
$r(\ell_H^-)\leq r(-\ell_H^-)\leq r(-\ell_H^+)\in [1,r_*^-]$,  where $(r(\pm\ell_H^-),r(-\ell_H^+))$ are the orbits relative to the specific angular momenta $\{\pm\ell_H^-,-\ell_H^+\}$. It is clear  there exist two ranges of \textbf{BH} spin for $a<a_*$ and $a>a_*$, where
$a_*=0.875473$. Fig.\il(\ref{Fig:Plotqmminatint})  shows  also  the replica locations with respect to the \textbf{BH}  outer horizon  for different spins.
The orbits $r$ are, independently of the plane, close to the  \textbf{BH}   for $r\lesssim3.8$.
\subsection{{Celestial coordinates  $(\alpha,\beta)$ for the horizons replicas}}\label{Sec:shadows-replicas}
In this section, we analyze the $(\alpha,\beta)$ coordinates, solutions of the equations $(\mathfrak{R})$, for $\ell=\{\pm\ell_H^-,-\ell_H^+\}$.

Solutions of  $(\mathfrak{R})$ for $\ell=\{\pm\ell_H^-,-\ell_H^+\}$  provide the celestial coordinates $(\beta,\alpha)$  on the luminous edges, related to the fixed constraints   in accordance with
spin  $a$, radius $r$, angle $\sigma$ and Carter constant $q$. (Quantities $(a,r,\sigma,q)$  have been  discussed in Sec.\il(\ref{Sec:first-lun}).).

In our analysis,  we solve the equations $(\mathfrak{R})$ (providing  the shadow boundary), for null geodesics with $\ell=\pm\ell^\pm_H$ (replicas). Hence we individuate parts of the luminous edges generated by the null geodesics which are \emph{also} replicas, i.e.  we consider replicas that can form (and therefore can be observed from)  the \textbf{BH}  shadow boundary, as certain angle $\theta$ and  for certain spin $a$. Replicas  location on the shadow profiles are then studied  for all values of  $\theta$ and $a$.

\begin{figure}[h]
\centering
\includegraphics[width=8cm]{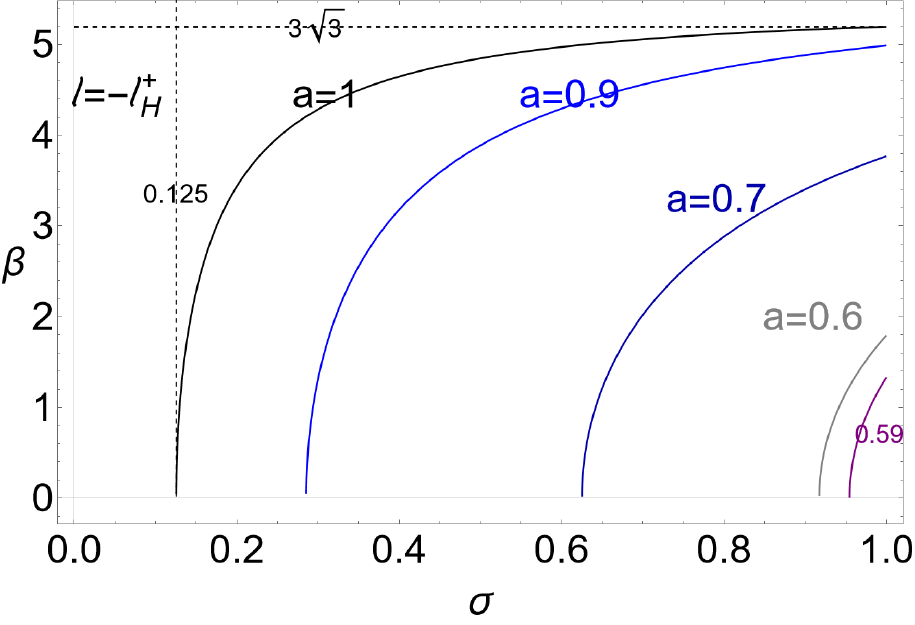}
\includegraphics[width=8cm]{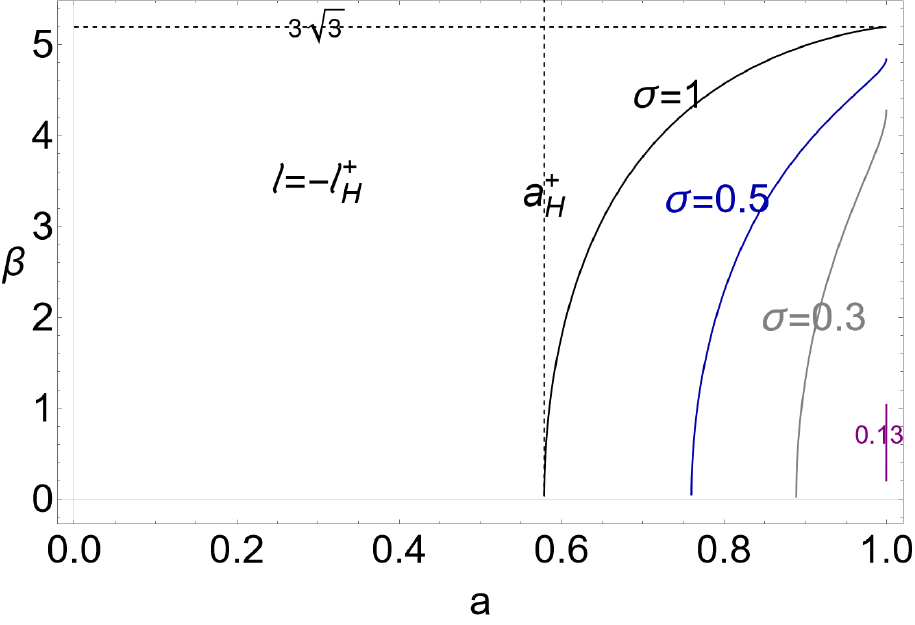}
\includegraphics[width=8cm]{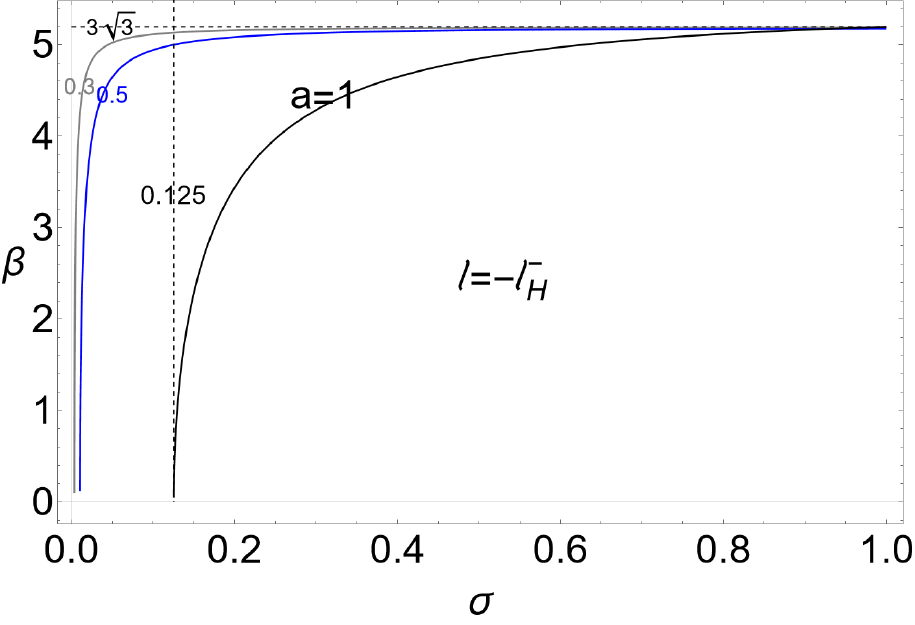}
\includegraphics[width=8cm]{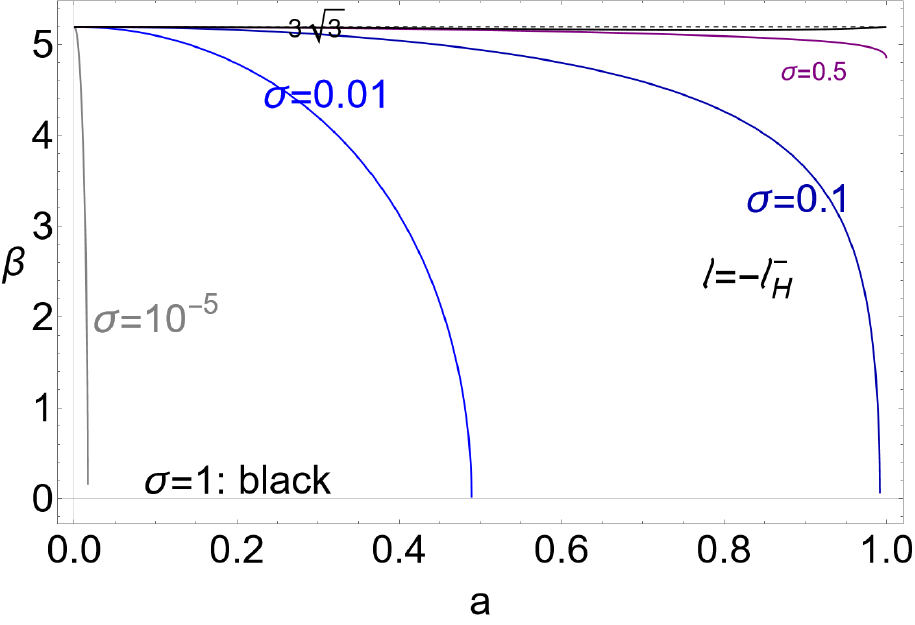}
\includegraphics[width=8cm]{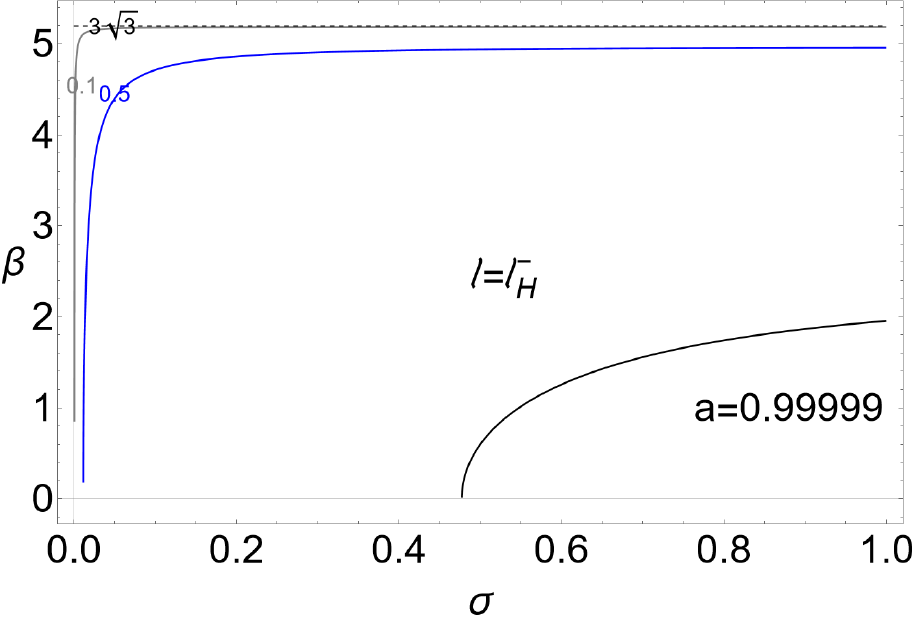}
\includegraphics[width=8cm]{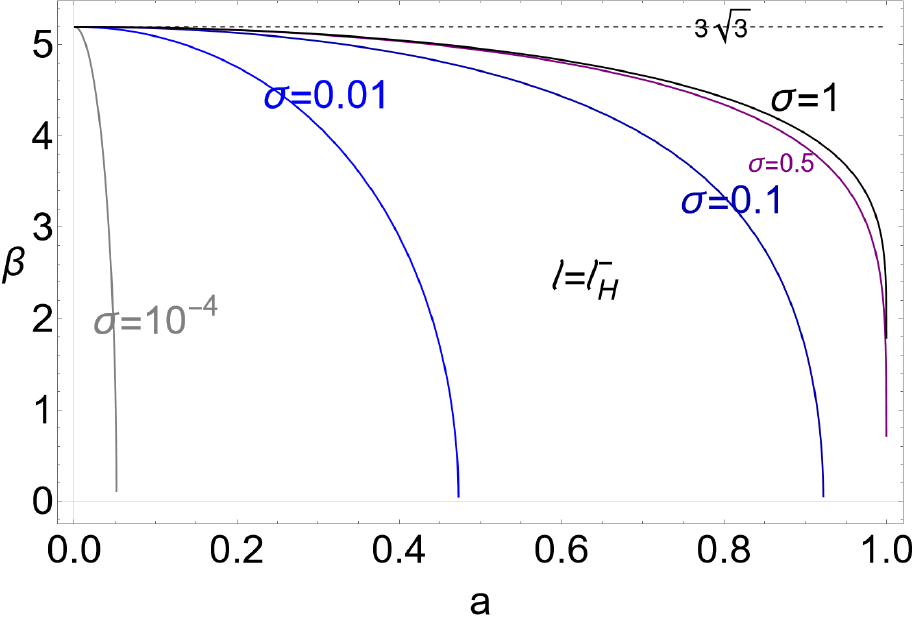}
\caption{The celestial coordinate $\beta$, {solution of the set $(\mathfrak{R})$ of Eq.\il(\ref{Eq:radial-condition})  with the constraint}  $\ell=-\ell_H^+$ (upper panels), $\ell=-\ell_H^-$ (center panels) and  $\ell=\ell_H^-$ (bottom panels)
	($\ell_H^\pm$ is the angular momentum of the outer and inner  horizon respectively)
	 as a function of the angular coordinate $\sigma\equiv \sin^2\theta\in[0,1]$ for different values of the spin $a$ (left panels), and in terms of the spin $a$ for different values of $\sigma$ (right panels).
	    The dotted lines are the limiting values of $\sigma$, $\beta$, and $a$. All the quantities are dimensionless.}\label{Fig:Plotbetrodon}
\end{figure}

\begin{figure}
\centering
\includegraphics[width=5.6cm]{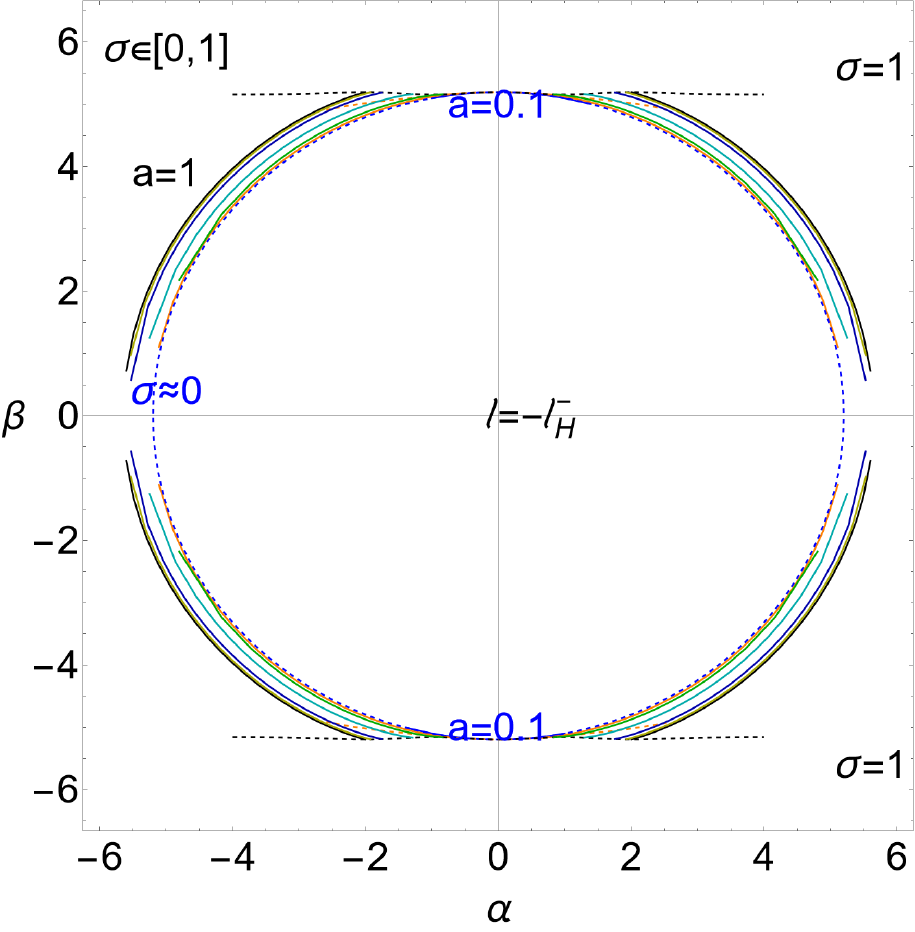}
\includegraphics[width=5.6cm]{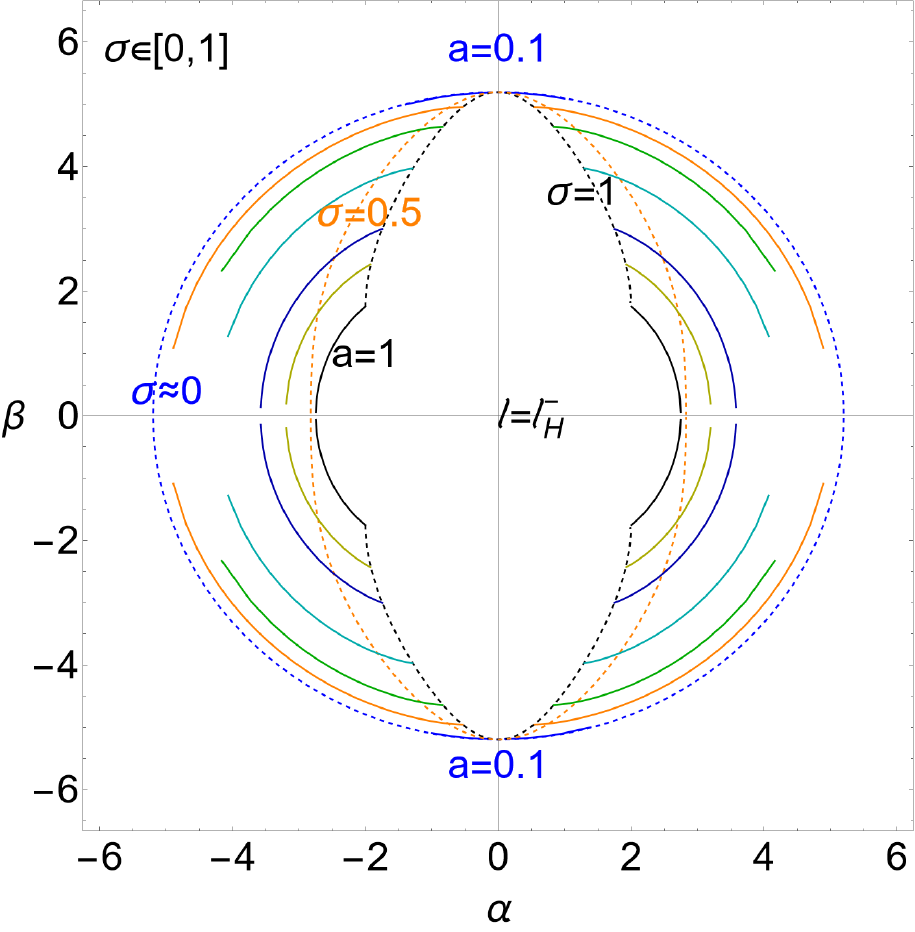}
\includegraphics[width=5.6cm]
{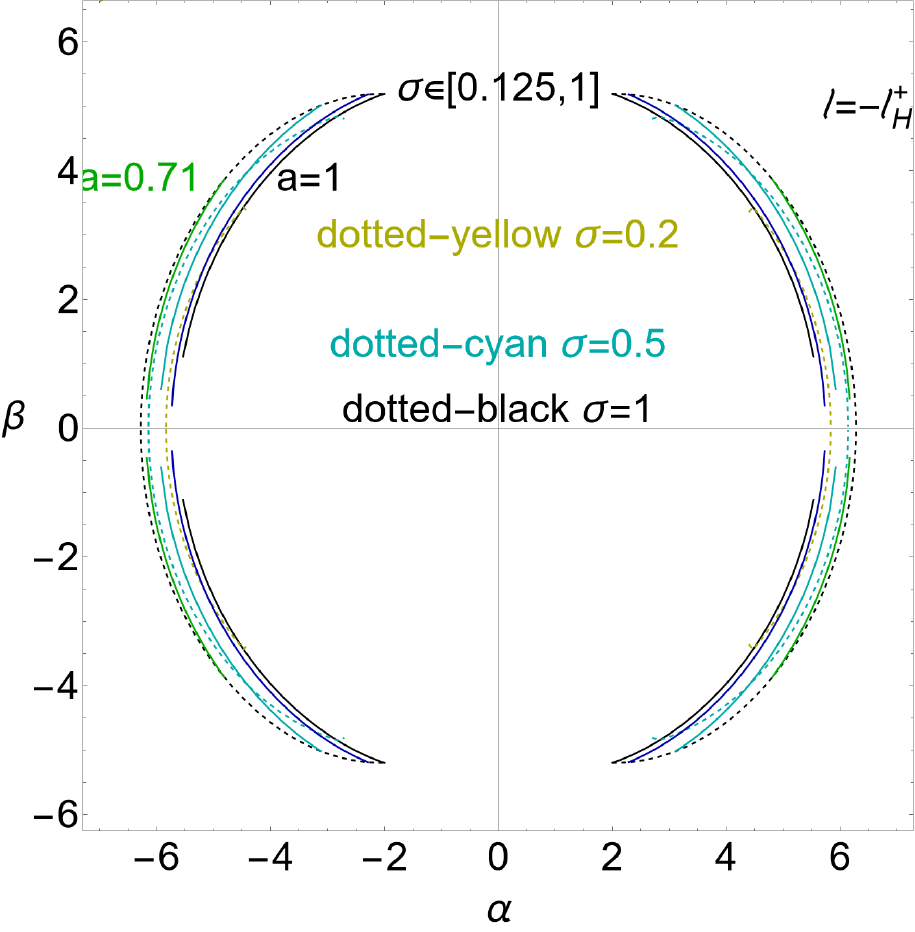}
\caption{{Solutions of the set $(\mathfrak{R})$ of Eq.\il(\ref{Eq:radial-condition})  (celestial coordinate $\beta$ as function of $\alpha$) with the constraint} $\ell=-\ell_H^-$ (counter-rotating inner horizons replicas--left panel), for $\ell=\ell_H^-$ (co-rotating inner horizons replicas--center panel) and for $\ell=-\ell_H^+$ (counter-rotating outer horizons replicas--right panel)  for different \textbf{BH} dimensionless spin $a\in[0,1]$ (solid curves) and angles $\sigma\equiv\sin^2\theta\in[0,1]$. The quantities $\ell_H^\pm$ are the outer and inner horizons angular momenta, respectively.  The black solid  curve is $a=1$ (outer curve in the left panel and inner curve in the center panel), the orange  solid curve is $a=0.5$, and the blue solid curve is for $a=0.1$ (outer curve in the right panel and inner curve in the center panel). Each solid curve is the set of replicas on the \textbf{BHs} shadows profiles at fixed \textbf{BH} spin $a$. Each point of a  solid curve is a replica for a fixed  plane $\sigma\in [0,1]$.
Each dotted curve is the set of  replicas on the \textbf{BHs} shadows profiles for a fixed plane $\sigma$. Each  point on a dotted curve is for a specific \textbf{BH} spin. Hence, the crossing of the dotted and solid lines,  fixes $\sigma$ and the spin $a$.
 For the co-rotating  inner horizons replicas (right panel),   the inner solid  curve corresponds to the   extreme  \textbf{BH} spacetime ($a=1$). The replicas  moves outwardly (in the plane $(\alpha,\beta)$) decreasing the \textbf{BH} spin. Viceversa,  in the counter-rotating case (counter-rotating  inner horizons replicas--left panel)  the extreme Kerr \textbf{BH} corresponds to   the outermost  curve of the plane,  and the curves move inwardly decreasing the \textbf{BH} spin.  Black dotted curves  correspond to  $\sigma=1$, blue dotted curves correspond to  $\sigma\approx0$. (In same cases, for  graphical reasons, dotted curves have been extended beyond the intersections with the  solid curves  for fixed $a\in[0,1]$). On the left panel, $\sigma=0$ is the inner limiting closed surface for the replicas in the shadow profile, while  the equatorial plane, $\sigma=1$, is the open liming outer curve (confinement on the equatorial plane). On the center panel the outer closed  limiting dashed blue  curve corresponds to $\sigma\approx0$. Right panel: the inner black solid curve is for $a=1$ and the green outer curve for $a=0.71$. All quantities are dimensionless.}\label{Fig:Plotverini1b}
\end{figure}

\begin{figure}[h]
\centering
\includegraphics[width=5.5cm]{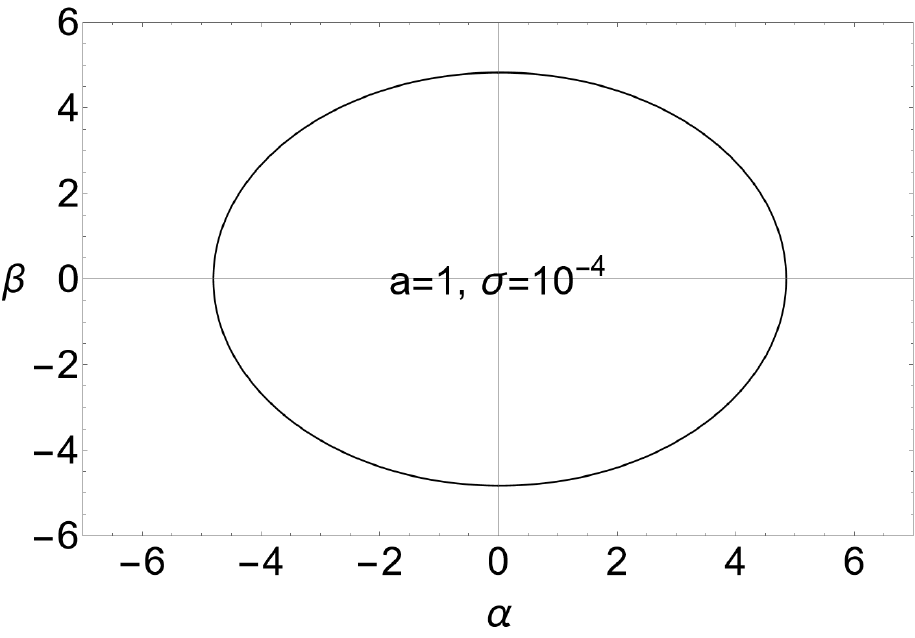}
\includegraphics[width=5.5cm]{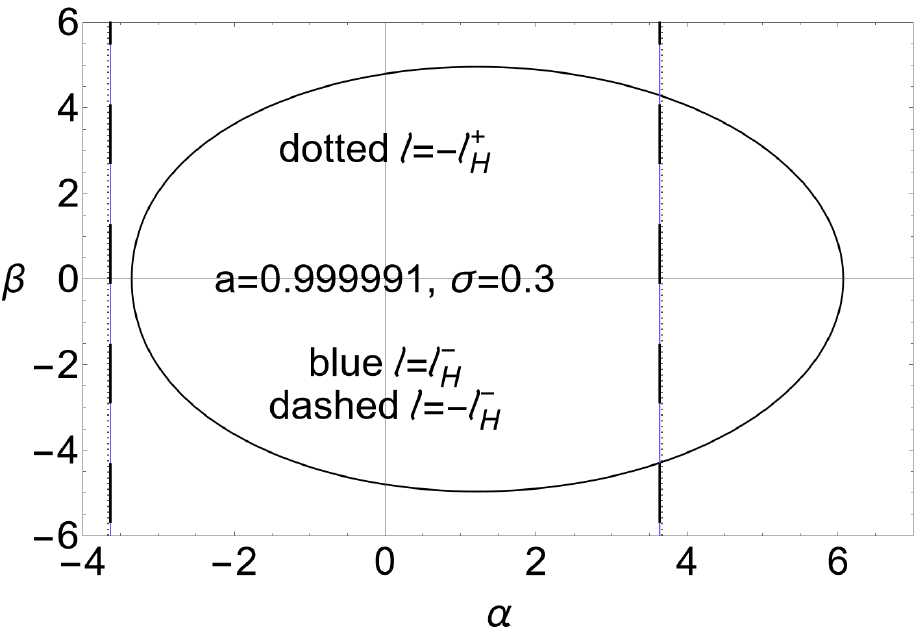}
\includegraphics[width=5.5cm]{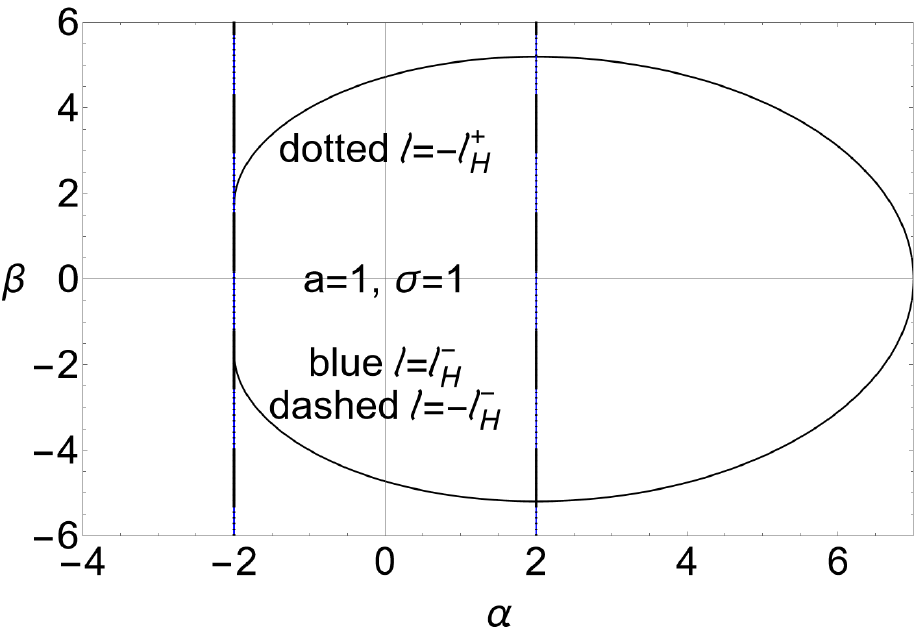}
\includegraphics[width=5.5cm]{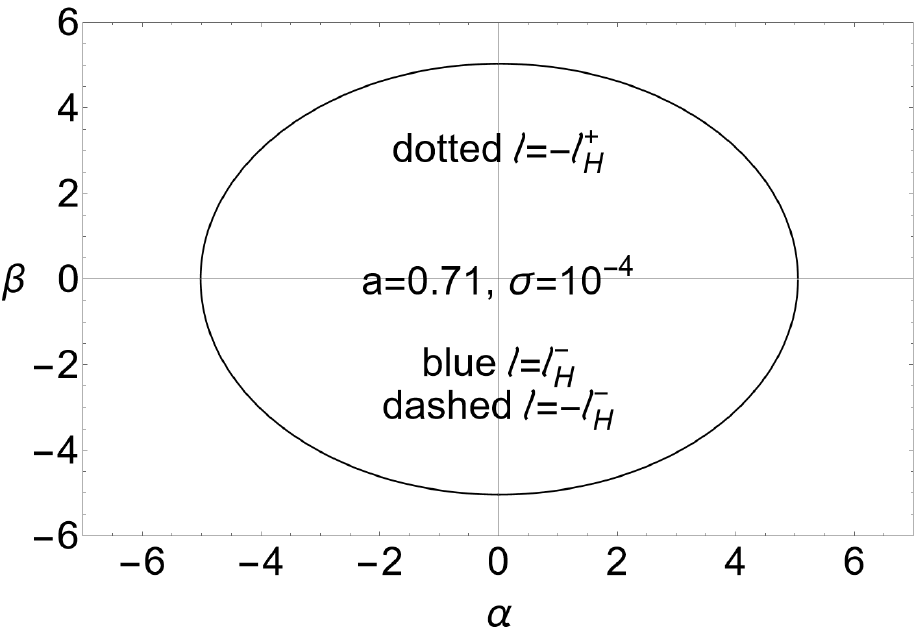}
\includegraphics[width=5.5cm]{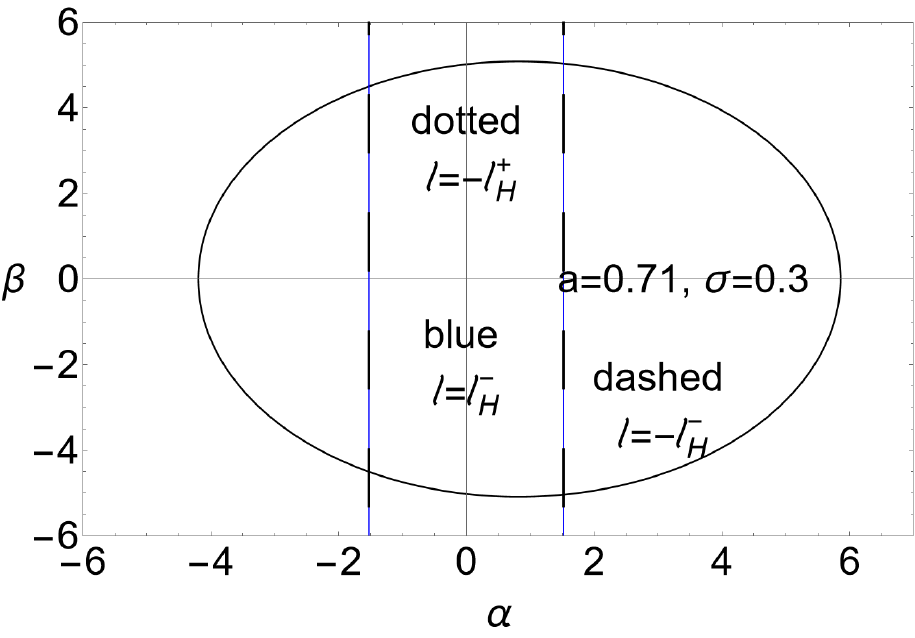}
\includegraphics[width=5.5cm]{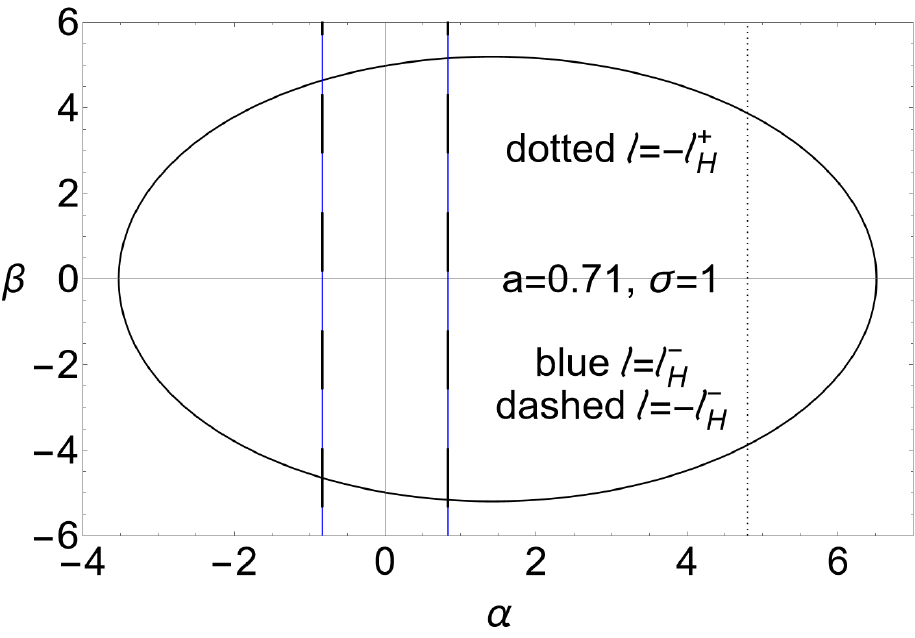}
\includegraphics[width=5.5cm]{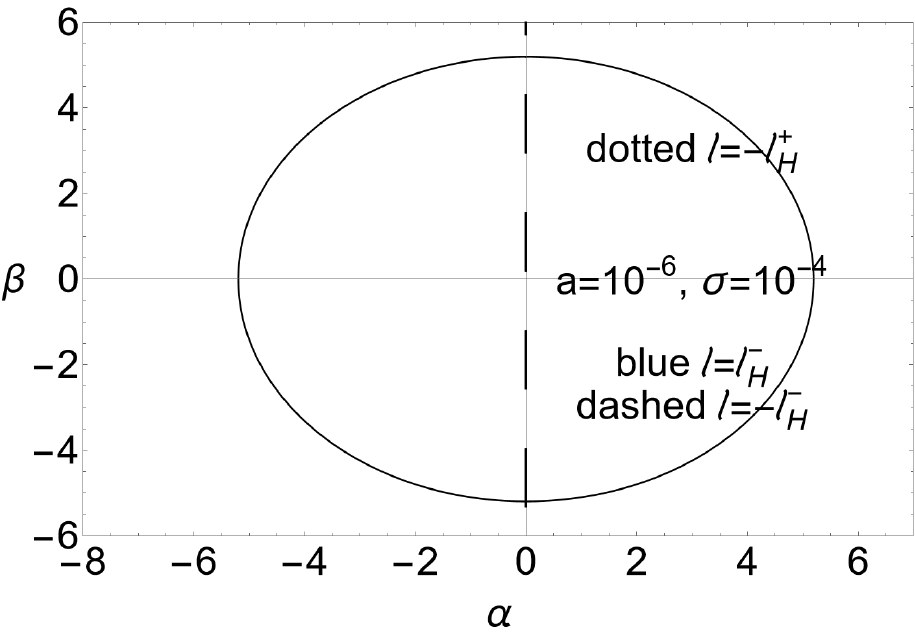}
\includegraphics[width=5.5cm]{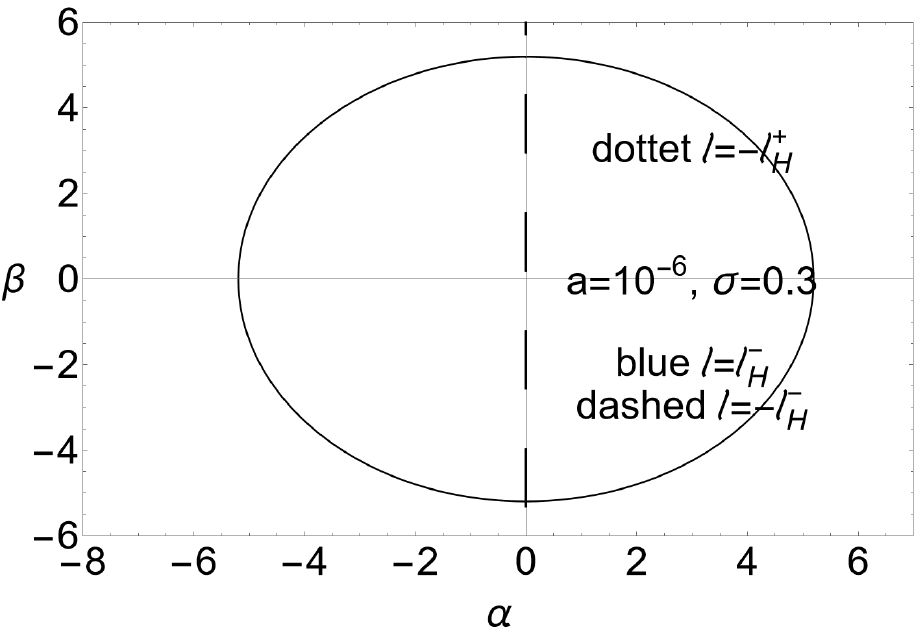}
\includegraphics[width=5.5cm]{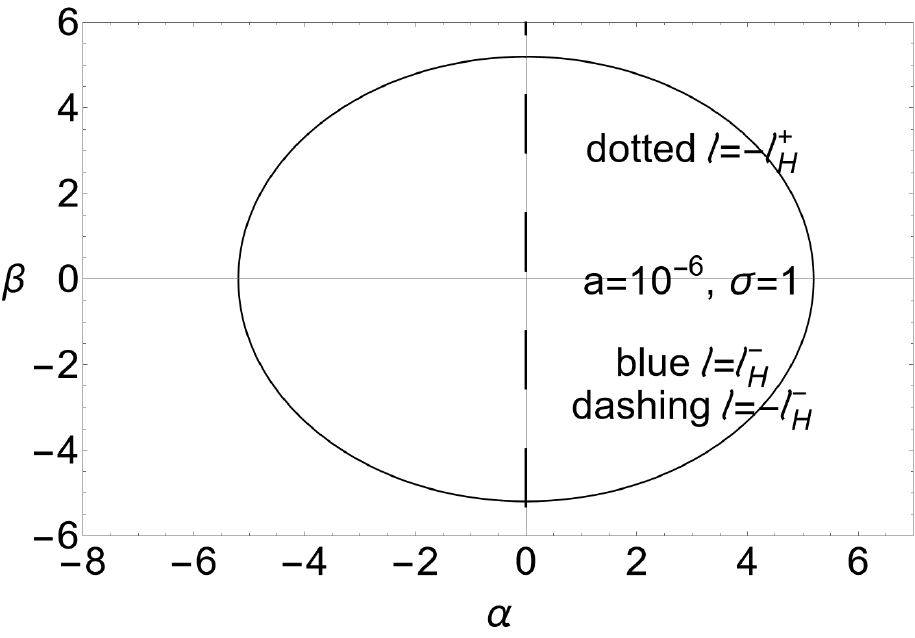}
\caption{\textbf{BH} shadow profiles (closed black curves) solutions of the set $(\mathfrak{R})$ of Eq.\il(\ref{Eq:radial-condition})      (celestial coordinate $\beta$ versus $\alpha$), for \emph{all} values of the impact parameter $\ell$, and  for different dimensionless  spins $a\in[0,1]$ and angular coordinate $\sigma\equiv \sin^2\theta\in[0,1]$, denoted on the panels.
Vertical lines on the panels are solutions of the set $(\mathfrak{R})$  (celestial coordinate $\beta$ versus $\alpha$), for $\ell=\pm\ell_H^\pm$ ($\ell_H^\pm$ are the outer and inner  horizons angular momenta) and for fixed  $a$ and  $\sigma$.  The upper row is for $a\approx 1$,  middle row for $a=0.71$, bottom row for $a\approx 0$, left column for $\sigma \approx 0$,  center column for $\sigma=0.3$, and right column for $\sigma=1$.  (All quantities are dimensionless).}\label{Fig:Plotzing1}
\end{figure}

Solutions of this analysis are shown in Figs\il(\ref{Fig:Plotbetrodon},\ref{Fig:Plotverini1b})  and Figs\il(\ref{Fig:Plotzing1}).
(Celestial coordinate $\alpha\equiv-\ell/\sqrt{\sigma}$  of Eq.\il(\ref{Eq:alpha-def}) as  function of the \textbf{BH} spin  on the equatorial plane and $\ell\in\{\pm\ell_H^-,-\ell_H^+\}$ is shown in Figs\il(\ref{Fig:Plotqmminatint})--right panel).

\medskip

More precisely:

\medskip

--In Figs\il(\ref{Fig:Plotbetrodon}), the
 celestial coordinate $\beta$, solution of the set $(\mathfrak{R})$ of Eq.\il(\ref{Eq:radial-condition})  with the constraint  $\ell=-\ell_H^+$ (upper panels), $\ell=-\ell_H^-$ (center panels) and  $\ell=\ell_H^-$ (bottom panels), is
	 shown
	 as a function of the angular coordinate $\sigma$ for different values of the spin $a$ signed on the curves (left panels), and in terms of the spin $a$ for different values of $\sigma$ signed on the curves (right panels).
	    The dotted lines are the limiting values of $\sigma$, $\beta$, and $a$.
	
\medskip	
	
{--In Figs\il(\ref{Fig:Plotverini1b}),  the celestial coordinate $\beta$, solution of the set $(\mathfrak{R})$ of Eq.\il(\ref{Eq:radial-condition})  with the constraint $\ell=-\ell_H^-$ (left panel), for $\ell=\ell_H^-$ (center panel) and for $\ell=-\ell_H^+$ (right panel),  is  shown as function of $\alpha$  for different \textbf{BH} dimensionless spin $a\in[0,1]$ (solid curves) and angles $\sigma$.  In these panels,  each curve  corresponds to a \emph{fixed}  spin and a specific  constraint, and  each point of each  curve is for \emph{fixed} $\sigma$. Each solid curve is the set of replicas on the \textbf{BHs} shadows profiles at fixed \textbf{BH} spin $a$. Each point of a  solid curve is a replica for a fixed  angle $\sigma$.
Each dotted curve is the set of  replicas on the \textbf{BHs} shadows profiles for a fixed plane $\sigma$. Each  point on a dotted curve is for a specific \textbf{BH} spin. Hence, the crossing of the dotted and solid lines,  fixes $\sigma$ and the spin $a$. Hence, each point   $(alpha,\beta)$ of a curve provides the corresponding  solutions on the \textbf{BH} shadow profile.}

\medskip

{---In Figs\il(\ref{Fig:Plotzing1})    \textbf{BH} shadow profiles   (celestial coordinate $\beta$ versus $\alpha$), solutions of the set $(\mathfrak{R})$ of Eq.\il(\ref{Eq:radial-condition}) for all values  $\ell$,  are shown  for different dimensionless  spins $a\in[0,1]$ and angular coordinate $\sigma$ (closed black curves). Points corresponding to  null geodesic solutions of the set $(\mathfrak{R})$ within constraints $\ell=\pm\ell_H^\pm$  are vertical lines on the panels.
Solving $(\mathfrak{R})$ with constraints on $\ell$,  and \textit{fixing} $a$ \emph{and} $\sigma$,  we obtain  a set of points $(\alpha,\beta)$ on the shadow profile for the fixed $(a,\sigma)$,  as shown  in Figs\il(\ref{Fig:Plotzing1}) (closed curves of Figs\il(\ref{Fig:Plotzing1}) obtained at fixed $(a,\sigma)$ for \emph{all} values of $\ell$.}

{Varying $a$  and $\sigma$,
we obtained in  Figs\il(\ref{Fig:Plotbetrodon},\ref{Fig:Plotverini1b})  a map of the constrained regions on the shadow profiles, for \emph{all} the observational angles $\sigma$ and for \emph{all} \textbf{BH} spins $a$.}

Solving $(\mathfrak{R})$ with constraints on $\ell$ at \emph{fixed} $a$ and for \emph{all} values of the observational angle $\sigma$,  provides a set of points on the shadow profile representing the superimposition of  the points on the shadow  boundary relative to the constrained null geodesics,  for \emph{all} values of the observational angle.
 These sets of points  are therefore  curves of the $(\alpha,\beta)$ plane
for different spins.

\medskip

In details, the coordinate  $\beta$ is independent from the rotation orientation (sign of $\ell$) and is  upper bounded by the value $\beta^2<27$.

{Let us concentrate on the
\textbf{BH}  counter-rotating (with $\ell=-\ell_H^-$) and  the co--rotating ($\ell=\ell_H^-$)  inner horizons replicas on the shadow profile shown in Figs\il(\ref{Fig:Plotverini1b})  for different  spins $a\in[0,1]$ (solid curves) and angles $\sigma\in[0,1]$. The black solid  curve (outer curve in the left panel and inner curve in the center panel) represents the situation  for the extreme Kerr \textbf{BH} ($a=1$), and the blue solid curve (outer curve in the center panel and inner curve in the left panel) is for $a=0.1$.  (In other words,  each solid curve represents the set of replicas on the  shadows profiles for a fixed spin $a$. Each point of the solid curve is a replica for a \emph{fixed}    $\sigma\in [0,1]$.).
On the other hand, each dotted curve is the set of  replicas for a \emph{fixed} parameter  $\sigma$. Each  point of a dotted curve is for a specific \textbf{BH} spin.  Black dotted curves  correspond to  $\sigma=1$, blue dotted curves correspond to  $\sigma\approx0$. (In same cases, for  graphical reasons, dotted curves have been extended beyond the intersections with the  solid curves). On the left panel, $\sigma\approx 0$ is the inner limiting closed surface for the replicas in the shadow profile, while  the equatorial plane, $\sigma=1$, is the open limiting outer curve. On the center panel, the outer closed  limiting dashed blue  curve corresponds to $\sigma\approx0$.
 For the co-rotating  inner horizons replicas,   the inner solid  curve corresponds to the   extreme  \textbf{BH} spacetime ($a=1$). The curves corresponding to   replicas  move outwardly (in the plane $(\alpha,\beta)$) decreasing the value of the spin. Viceversa,  in the counter-rotating case (counter-rotating  inner horizons replicas),  the extreme Kerr \textbf{BH} corresponds to   the outermost  curve of the plane,  and the curves move inwardly with decreasing spin values.}
{In all cases, $\beta$ is upper bounded by the value $\beta=3\sqrt{3}$.
For $\ell=-\ell_H^+$, the value of $\beta$ increases with the spin and with  $\sigma$. There are no solutions for slowly spinning \textbf{BHs} and close to the \textbf{BH} poles.
For $\ell=-\ell_H^-$, in general,
$\beta$ decreases with the \textbf{BH} spin and increases with $\sigma$.
For $\ell=\ell_H^-$  (co--rotating inner horizon replicas) the  results are  similar  to the case
$\ell=-\ell_H^-$ ($\beta$ is even in $\ell$).}
%
%
%

\section{Discussion and concluding remarks}\label{Sec:conclu-Rem}
This work explored
the  connection between \textbf{BH} shadow profile and horizons  replicas,   analysing  the  solutions of the equations $\mathfrak{(R)}$ of Eq.\il(\ref{Eq:radial-condition})     for the \textbf{BH} shadow profile, assuming constant  $\ell$ coincident  in magnitude with the angular momentum of the \textbf{BH} horizon $\ell_H^\pm(a)$ (horizons replicas).
These  special solutions  of    $\mathfrak{(R)}$ are
points and regions on the \textbf{BH} shadow luminous profiles, here  highlighted and studied at different observational angles and for all values of the \textbf{BH} spin parameter.
Conditions under which horizon replicas are imprinted in the \textbf{BH}  shadows have been examined.

Replicas appear  on particular regions of  the orbital boundary defining the \textbf{BH} shadow.
In this analysis, the  $\beta$ function  is the  significant celestial coordinate  in  the determination of the   shadows topology and shape.
The results  are discussed in Sec.\il(\ref{Sec:lblu}) and in particular for the $(\alpha,\beta)$ coordinates in Sec.\il(\ref{Sec:shadows-replicas}).

\medskip

More in details:

\medskip

From a methodological view--point, we solved Eqs.\il(\ref{Eq:radial-condition})
(refereed as set $(\mathfrak{R})$) fixing $\ell=\pm\ell_H^\pm$.
{In general,
for fixed $a$ and $\sigma$, the \textbf{BH} shadow profile is found in terms of solutions of  the set $(\mathfrak{R})$  for \emph{all} values of $\ell$ and $r$. Solutions are, therefore, closed curves $\beta(\alpha)$ (black  curves of Figs\il(\ref{Fig:Plotzing1})).
Solving $(\mathfrak{R})$ with constraints on $\ell$,  and \emph{fixing} $(a,\sigma)$,  we obtain  a set of points $(\alpha,\beta)$ on the shadow profile corresponding to the fixed  \textbf{BH} spin $a$ and observational angle $\sigma$ (vertical lines  in Figs\il(\ref{Fig:Plotzing1})).}
On the other hand, solving $(\mathfrak{R})$ with constraints on $\ell$, at \emph{fixed} $a$ and for \emph{all} values of the observational angle $\sigma$,  provides a set of points on the shadow profile representing the superposition of  the points on the shadow  boundary relative to the constrained null geodesics,  for \emph{all} values of the observational angle.
 {These sets of points  are, therefore,  curves of the $(\alpha,\beta)$ plane, and they are shown, for example in  Figs\il(\ref{Fig:Plotverini1b})
for different spins.  In these panels,  each curve  corresponds to a \emph{fixed}  spin and a specific  constraint, and  each point of each  curve is for a different  $\sigma$.
 In this way, we obtained, for \emph{all} \textbf{BH} spins,  a map of the constrained regions on the shadow profiles, for \emph{all} the observational angles.}
We also related, for each  constraint, the quantities
 $(a,\sigma,r,\ell,q,\beta)$---
Fig.\il(\ref{Fig:Plotqmminatintsigma},\ref{Fig:Plotqmminatutt},\ref{Fig:Plotbetrodon}).
The quantities $(a,r)$  in the constrained solutions of $(\mathfrak{R})$   are related  as shown in
Figs\il(\ref{Fig:Plotqmminatint}).
The relations $(\sigma,r)$ and $(\sigma,a)$ from the solutions $(\mathfrak{R})$    with the constraints  are in
Figs\il(\ref{Fig:Plotqmminatintsigma}).
Thiese analysis set the range of values for the observational angle $\sigma$ and
for range of  spin $a$, where replicas appear on the \textbf{BH} shadow luminous boundary.
The quantities $(q,a)$ and $(q,r)$, constrained solutions of the set $(\mathfrak{R})$,    are in
Figs\il(\ref{Fig:Plotqmminatutt}).

The results of this analysis, illustrated in details  in   Sec.\il(\ref{Sec:first-lun}), show that, in general, the radius  $r$ is bounded  in a small range of values  (around $r=3$ for any spin and angle $\theta$)  and the angle $\sigma$  is, in general,  bottom bounded   according to  Figs\il(\ref{Fig:Plotqmminatint},\ref{Fig:Plotqmminatintsigma},\ref{Fig:Plotqmminatutt}).
The results are then confirmed from the analysis in  Figs\il(\ref{Fig:Plotbetrodon},\ref{Fig:Plotverini1b}).

\medskip

--In Figs\il(\ref{Fig:Plotbetrodon}),
 celestial coordinate $\beta$ (solution of the set $(\mathfrak{R})$  with   $\ell=-\ell_H^+$  and $\ell=\pm\ell_H^-$)
	is  shown
	 as a function of the angular coordinate $\sigma$ for different  spin $a$ signed on the curves (left panels), and in terms of the spin $a$ for different values of $\sigma$ signed on the curves (right panels).

{--In Figs\il(\ref{Fig:Plotverini1b}), the coordinates   $\beta$ and  $\alpha$, solutions of  $(\mathfrak{R})$ with the constraint $\ell=-\ell_H^-$ (left panel), for $\ell=\ell_H^-$ (center panel) and for $\ell=-\ell_H^+$ (right panel),  are  shown  for different \textbf{BH}  spins $a\in[0,1]$ (solid curves) and angles $\sigma$. Each point   $(\alpha,\beta)$ of a curve provides the corresponding solutions on the \textbf{BH} shadow profile. The crossing of the dotted and solid lines  fixes $\sigma$ and the spin $a$. Each solid curve  corresponds to    the set of replicas on the \textbf{BHs} shadows profiles at \emph{fixed} \textbf{BH} spin $a$ and a specific  constraint, and  each point of each  curve is   a replica for a fixed  angle $\sigma$, where
each dotted curve is the set of  replicas on the \textbf{BHs} shadows profiles for a fixed plane $\sigma$. (Each  point on a dotted curve is for a specific \textbf{BH} spin.).
Remarkably, the radial distance from the central attractor is $r<3.8$,  considering co-rotating and counter-rotating orbits from  any planes and for any spin, where   the celestial coordinate is upper bounded by the  value
$\beta<3\sqrt{3}$.
There are no solutions for the condition $\ell=\ell_H^+$ (co-rotating outer horizon  replicas).
It is clear that  the counter-rotating replicas of the outer horizons can be observed  far from the rotational axis for fast spinning black holes, i.e., for  $a\gtrsim0.59$ with  $\sigma\gtrsim0.125$.}

{--In Figs\il(\ref{Fig:Plotzing1}),    \textbf{BH} shadow profiles   (solutions of the set $(\mathfrak{R})$ of Eq.\il(\ref{Eq:radial-condition}) for \emph{all} values  $\ell$),    are shown  for different   spins $a\in[0,1]$ and  $\sigma$ (closed black curves). Vertical lines point to the solutions corresponding to  null geodesics  of the set $(\mathfrak{R})$ with the constraints $\ell=\pm\ell_H^\pm$.}

\medskip

 The constraints we set are therefore related to particular regions of the  orbital boundary (luminous edge) defining the  \textbf{BH} shadow. The imprints of replicas   (coordinates $(\alpha,\beta)$ solutions of constrained $\mathfrak{(R)}$) on the  shadow boundary of the central Kerr \textbf{BH}    map  the  luminous  edge at different observational angles, and for all \textbf{BH} spins. Distinguishing  in particular co-rotating and counter-rotating photons, for fixed plane $\sigma$ and spin $a$  (as shown in Figs\il(\ref{Fig:Plotzing1}))   and for \emph{all}  planes $\sigma$ and spin $a$.

The  existence and  location of replicas  on the shadow profiles vary according to  the value of the spin; for some spins there are no replicas.  This fact has  remarkably  consequences from an observational viewpoint as it  constitutes  a valid discriminant for the evaluation of the \textbf{BH} spin.
The new observables are  related to particular photon orbits, carrying information in their impact parameter about the angular momentum of the rotating \textbf{BH}, to potentially accompany other existing methods and  techniques to detect and measure the spins of   an accreiting \textbf{BH}. These methods  consist for example of
\textbf{X}--ray reflection spectroscopy  (see \cite{Reynolds:2020jwt}), or  gravitational waves detection  of
\textbf{LIGO-Virgo-KAGRA} (\textbf{LVK}) Collaboration\footnote{https://www.ligo.org/index.php; https://www.virgo-gw.eu;https://gwcenter.icrr.u-tokyo.ac.jp/en/}
  \cite{LIGOScientific:2021djp}.
 (We should  also stress that \textbf{BH} shadows can depend on properties   of the region of the  light
distribution  and its  source for example  accretion disks, where
photons could   interact with  the accreting  plasma. Our  analysis can be   adaptable  to and complement  the constraints imposed by the specific numerical or analytical model of accretion disks--see for example  \textbf{EHT} Collaboration  compared  numerical torus models
directly to observations in several comprehensive analysis \cite{EHT2e,EHT22e} see also \cite{EHT22f,EHT21b,EHT2,EHT2d,EHT2f,
White20,Porth,Janssen,Chatterjee20etaletal,
Emami21,Lucchini19,Vincent19,Curd23,
Anantua23,Gralla19,Gralla20,Johnson,
Narayan19}.).

Analysis of the photons parameters $\ell$  could  help understanding  the \textbf{BH} energetic   processes. (In particular the poles  of a Kerr \textbf{BH}.
 The quantity  $\ell$, the photons impact parameter  (specific angular momentum) adopted here to discern the   photons,   plays also a significant role in \textbf{BHs} accretion physics,  limiting the  \textbf{BH} energy extraction with the limiting quantity $\ell$ evaluated on the outer horizon.).
An interesting aspect to explore further is the relevance of the horizons replicas, observable  in  \textbf{BH} shadows, for the \textbf{BH} magnetosphere constraints.
 Light surfaces  (related  to the horizons replicas) are widely adopted  in the analysis of the \textbf{BH} magnetosphere--see for example \cite{Punsly,Camenzind})--, constraining the electromagnetic fields, which link, for example, the disk to the central \textbf{BH} attractor\footnote{
For an electromagnetic field
in Kerr spacetime with $\omega< \omega_H^+$,    light surfaces (\textbf{LSs}) are   surfaces  where the magnetic field lines (co)-rotate (with $\omega$) at the speed of light. ($\omega< \omega_H^+$  is the outer horizon tangency  condition in the bundle approach (see Sec.\il(\ref{Sec:MB-Appendix1})),   and it  constraints the  energy extraction from the \textbf{BH}. In the bundles framework, however, we consider also the counter-rotating condition  for \emph{fixed} $\omega<0$.).
 In the Kerr \textbf{BH} spacetime, there are two \textbf{LSs}.
 The inner one is  inside the ergoregion
 and one is  outside  the ergoregion \cite{Z77,BZ77,Pu15,UzS,TNM,Pan}, corresponding  to the  Pulsar light cylinder
asymptotically.
The
 \textbf{BH} magnetosphere, rigidly rotating with angular velocity $\omega$,  is divided into
 sub-luminal and super-luminal rotation regions, depending on
  $\mathcal{L}_{\mathcal{N}}\equiv \mathcal{L}\cdot\mathcal{L}$ sign ($\mathcal{L}\equiv \xi_{(t)} +\omega\xi_{(\phi)}$). The separating surfaces, solutions of  $\mathcal{L}_{\mathcal{N}}=0$, are \textbf{LSs}
\cite{Komissarov,Uz05,Mahlmann18}.
Thus, outside of the outer \textbf{LS}, magnetic
field lines  would rotate faster than the speed of light (respect
to ZAMOs). Inside the inner \textbf{LS},  magnetic
field lines would  (counter)rotate super--luminally with respect to the
ZAMO \cite{Z77,BZ77,Komissarov,Macdonald,KMCK,McKinney,Crinquand,CKF,CKP,MCN07,NC14}
In the  Grad--Shafranov (GS) equation (from the  poloidal component of the force-balance equation) for the Kerr magnetosphere,  $\omega$ is the velocity of the magnetic field lines.
In the \textbf{BH} magnetospheres,  in the  force-free limit,   the \textbf{LSs} correspond to two singular surfaces   (together with the event horizon, which coincides with the singular surfaces of the Alfven waves-the relativistic
 generalization of the  Alfven points  in   force-free conditions).
On \textbf{LSs},   regularity conditions  can be  set  to determine the poloidal current as a function of the poloidal
magnetic flux.
 The event-horizon
regularity condition (Znajek  horizon boundary condition) can be used to  determine the poloidal flux
distribution on the horizon.
The GS equation is
a second-order differential equation for the magnetic flux
with two eigenfunctions $I$ and $\omega$ to be
determined, for example,  by requiring that the magnetic
field lines smoothly cross the \textbf{LSs}, where the
GS  is a  first-order differential equation. Therefore, the \textbf{LSs} are found by imposing that the  second-order derivatives of the GS equation   vanish there.
The solution depends on the distributions of the
magnetic field lines angular velocity
$\omega(\psi)$ and the poloidal electric current $I(\psi)$,  which can be  determined
self-consistently  by assuming  that the magnetic field lines cross  the   inner and outer \textbf{LSs}.
Clearly,  the event horizon corresponds to the solution of
the singularity condition for $\omega = \omega_{H}^+$,  explained in the bundles frame.
}.
The consequences of the existence of  horizons replicas on the  \textbf{BH}  magnetosphere will be considered  in future analysis.

\medskip

However   results found in this analysis    could constitute a new template of observations (and    a way to  identify the central attractor constraining the \textbf{BH}  dimensionless spin) in future observational  enhancements.

More specifically, we  should consider that the current  \textbf{EHT} images consist of
unresolved rings
composed by    sets of (infinite)  narrow self--similar sub--rings (photon orbits), which are usually   indexed by  the photon orbits numbers $n\in \mathrm{N}$  around
the \textbf{BH} (for  photons
circling  the attractor $n/2$ times before reaching the observer), approaching  and converging, asymptotically,  the \textbf{BH} shadow boundary\footnote{\textbf{EHT} observations of the horizon
scale emission of \textbf{M87*} could support the presence of a "narrow
ring--like" feature \cite{Avery}.
On the other hand, evidence for the
presence of a lensed photon ring in the 2017 \textbf{M87*} observations was discussed in\cite{2022MNRAS.517.2462L}.
It has been
argued that  the  photon ring (simultaneously with the  "inner shadow") could  be  visible in
sub--millimeter images of \textbf{M87*}  from \textbf{VLBI}, and the photon ring morphology  and
inner shadow can be used  to provide a \textbf{BH}  mass and spin estimation--\cite{918}.
The structure of emission
inside the critical curve has been considered for equatorial-disk emission models--see also \cite{918,2023ApJ...944...55P}.
A discussion on the possibility that multiple measurements
of the photon rings of \textbf{M87*},  could  resolve the  ring size and guess the emission region, is presented in \cite{Avery2}. There,  it is also claimed that
mass and spin may be guessed by the  $n = 2$ photon ring.}.
The critical curve defines  the region of light rays  captured by the photon shell and
(unstably) orbiting  the \textbf{BH}, falling
into the \textbf{BH}, or escaping to infinity.
The \textbf{BH} shadow itself is  the  interior of the critical curve.
The number of
oscillations of the light rays,
very near the critical curve     diverges
logarithmically as the coordinate approaches a
point on the critical curve.
Each point of the curve, parametrized  with the
$(\alpha,\beta)$ coordinates, is  associated with a null
geodesic  labeled by the  two
conserved quantities ($\ell,q$) which, as done  in our analysis,  in turn can be retraced to infer information on the  \textbf{BH} properties.

On the other hand,  resolved  \textbf{EHT} images   in  the  sub--rings,
could  provide  a method to determine the \textbf{BH} parameters by  evaluating for example the  sub--rings  morphological characteristics as  size, shape, or  thickness\footnote{
For example, the measurement of the  \textbf{M87} \textbf{BH} spin from its
observed twisted light was considered in
\cite{Tamburini}
(by the  characterization  of the electromagnetic
waves  vorticity from the \textbf{BH} surroundings).} that use the ring shape,
sub--rings and a variety of
models considered for the source.


The main morphological photon  ring characteristics, however,   are   below the current  \textbf{EHT} imaging resolution.
The
present   (as for \textbf{M87*}) array coverage
  does not allow  the
extraction of  dim features below $\approx 10\%$ of the peak brightness, i.e., it is estimated that  the lensed images account for $\approx 10\%$ of the flux (\textbf{GRMHD} simulations  suggest that
the photon ring should provide only $\approx 10\%$ of the total image flux
density, hence  precluding to the current  observations  the determination of
photon ring and  the resolution of its substructure)\footnote{
 The critical
curve analysis from direct observation  would imply   the detection   of  lensed sub--rings,  possible only with  a very--high resolution baselines. It remains possible to eventually \emph{infer} the features of the critical curve by implementing adapted method to the resolution of the
 sub--rings. A set of astrophysical models has been analysed and different  resolution techniques   have been implemented.
Furthermore, the ring  can  exhibit additional substructure due to the presence of not uniform light source as for example   an accretion disk.
 Indeed, it  should also be considered that the analysis is complicated by the fact  that, the size, shape, and depth of the observed brightness depression depend on the peculiarities  of  the emission process and emission region. (\textbf{GRMHD} simulations are primary
 tools to  resolve the features of the
 \textbf{EHT} images,  which are currently unable to structure the inner shadow  due to  lack  of  resolution and dynamic range, hopefully reachable in the next--generation \textbf{EHT}.)
Therefore, the bundle of  photon rings, substructuring  the final image, encodes   both the emission (governed by the orbiting matter and emission processes)
and the spacetime properties (determined by the \textbf{BH} only).}.

However, we should consider that  the inference of
the underlying image of the \textbf{EHT} observation  features is  probabilistic, using a variety of possible analytical models and   image
reconstruction based on  geometric modelling. Imaging  algorithms provide  a best fitting  image--\cite{Medeiros}.
 The characteristics considered in this analysis  should be  processed with  observations  constrained by the current resolution limits of \textbf{EHT} and also by the different  data analysis  implemented.  In the more straightforward scenario,  this  would  require
  distinguishing lensed sub--rings within the
observed image
with limited-resolution observations where, for example, the  systematic uncertainties may be  reduced using an enhanced ground or space-based
array.
Furthermore, the restriction of the  dynamic range and visibility
coverage  could be addressed by appropriate  modeling and imaging  able to disentangle   the  ring--like emission
structure, for example,
focusing on  multiple image reconstruction algorithms--\cite{2022ApJ...940..182T}.

Hence, until such limitations will be reduced any  efforts to use   more detailed
edge  properties
will be considered for future implementation or  enhanced technics (possible sub--rings  resolution methods in \textbf{EHT} extensions are discussed in \cite{Johnson},
 arguing on  the possibility of measurements   photon sub--ring from an  high--frequency ground array or low Earth orbits,
and  from Moon station.)
The results  we prove here  may be used in the frame of  future
observations by a  next-generation \textbf{EHT}, with proposed future Earth--based arrays and possibility of space--based stations, with their associated long
baselines which could  resolve first orders  photon rings. Indeed, in coming years more radio telescopes will be added to the
\textbf{EHT} array, resulting in   a larger and more sensitive array and increasing  the dynamic range of the current \textbf{EHT} of several factors,  higher--frequency observations will be possible,   reaching
higher signal--to--noise ratios and much better baseline coverage, obtaining an overall  improved
sensitivity.
%
\section*{Acknowledgements}
The work of HQ was supported by UNAM PASPA-DGAPA.

\appendix

\section{Notes on metric bundles and characteristic frequencies}\label{Sec:MB-Appendix}
In this section we provide more details on the concepts of
metric bundles  and replicas  used  to constrain solutions of system $(\mathfrak{R})$. In Sec.\il(\ref{Sec:MB-Appendix1}) the concept of metric bundles and replicas is  deepen.
Relations between definition of  photon circular  orbits  and  replicas are discussed in Sec.\il(\ref{Sec:photon}).
 Sec.\il(\ref{Sec:photon-shere}) clarifies  the relation between  concepts of  photons spheres   and replicas. In Sec.\il(\ref{Sec:photonu}) we relate circular photons orbits and replicas.
\subsection{Metric bundles  and replicas}\label{Sec:MB-Appendix1}
In this section we introduce the concept of metric Killing bundles (\textbf{MBs}) used  to constrain solutions of system $(\mathfrak{R})$. \textbf{MBs}, extensively studied in  \cite{bundle-EPJC-complete,remnants,Pugliese:2022vni,Pugliese:2022xry,Pugliese:2021aeb,Pugliese:2021hfl,Pugliese:2020azr},
 are defined as the collections of light--like circular orbits with the same  constant orbital  frequency $\omega$, which is called \emph{characteristic bundle frequency}.
 This frequency results to be   also the frequency of a   \textbf{BH} horizon of the corresponding collection. %

\medskip

\textbf{Frequencies $\omega_H^\pm$ }

We start  by introducing the quantities $\omega_H^\mp$ (the inner and outer horizon  frequencies respectively) from the definition of \textbf{BH} Killing horizons.

The event horizons  of a spinning \textbf{BH}  are   Killing horizons   with respect to  the Killing field
$\mathcal{L}_H^\pm=\xi_{(t)} +\omega_H^{\pm} \xi_{(\phi)}$.

Defining the more general Killing field
$\mathcal{L}_\omega\equiv \xi_{(t)} +\omega\xi_{(\phi)}$, the quantity  $\mathbf{\mathcal{L_N}}\equiv\mathcal{L}_\omega\cdot\mathbf{\mathcal{L}}_\omega$ becomes null  for photon-like
particles with rotational frequencies $\omega_{\pm}$. We define also the \emph{angular momentum}  $\ell\equiv 1/\omega_\pm$.

Therefore, \textbf{MBs} correspond to the set of all  solutions of the condition
$\mathbf{\mathcal{L_N}}=0$ for a \emph{fixed} $\omega$ ($\ell$).

\begin{figure}
\centering
\includegraphics[width=9cm]{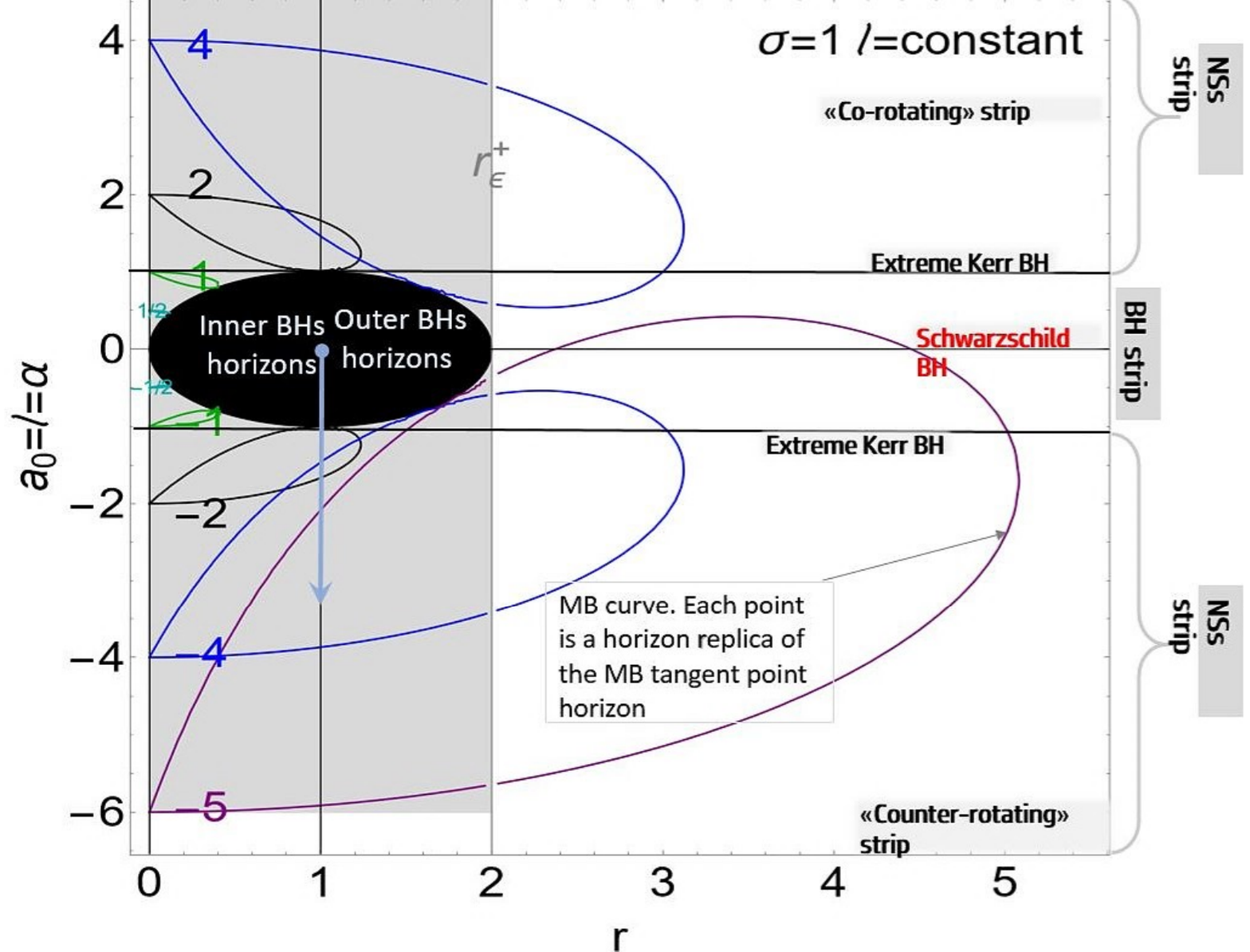}
\includegraphics[width=8cm]{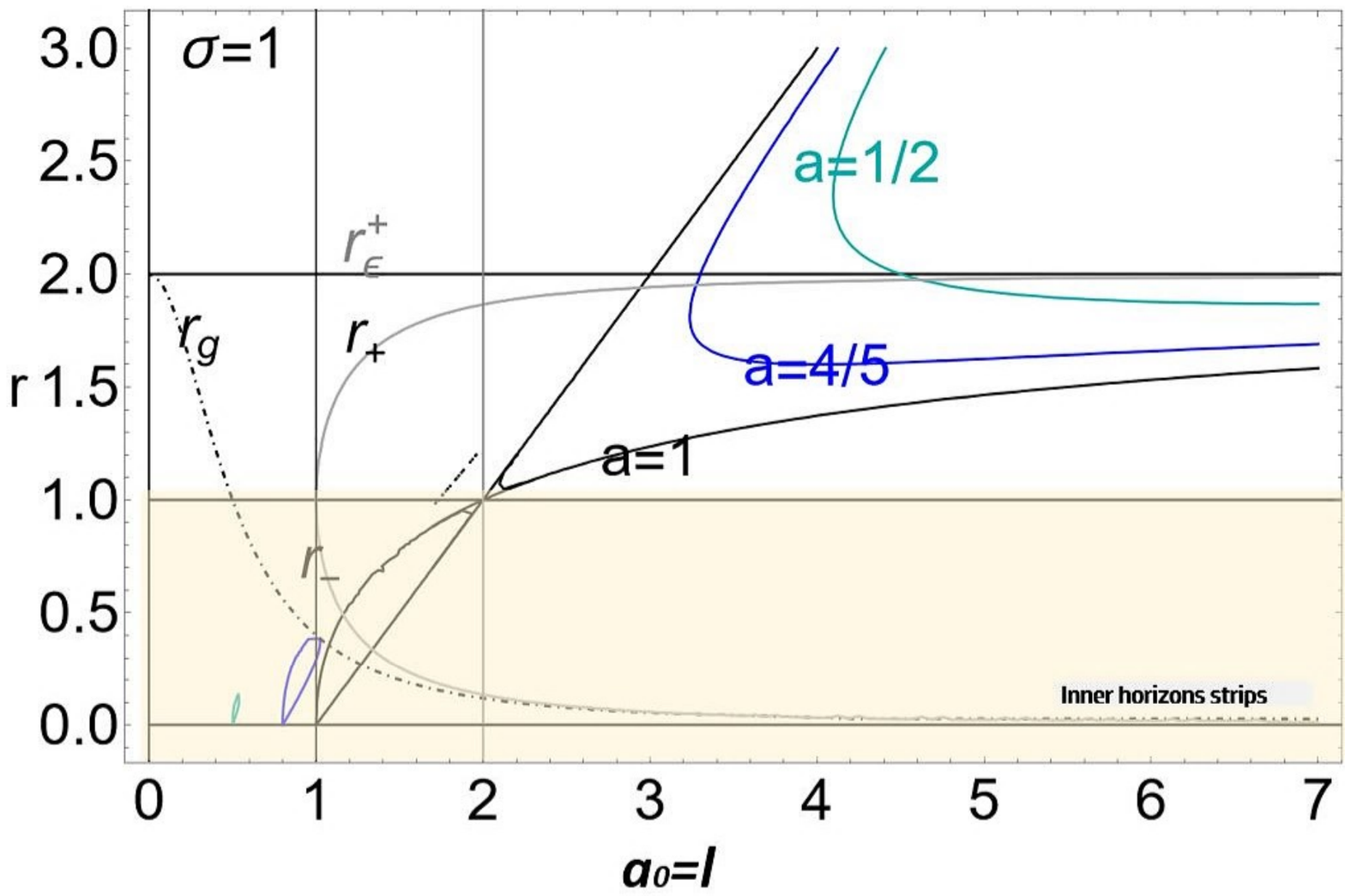}
\caption{Left panel: Extended plane $a-r$  on the equatorial plane $(\sigma=\sin^2\theta=1)$ of the Kerr geometry.  Metric bundles (\textbf{MBs})  with   characteristic frequency $\omega$ and angular momentum $\ell=1/\omega=a_0=$constant  are shown  for different values of $\ell$, which are denoted on each curve. The quantity $a_0$ represents the bundle origin spin at $r=0$, which coincides with  celestial coordinate $\alpha$--Eq.\il(\ref{Eq:alpha-def}). The \textbf{BH} region is shown in black  and the outer and inner horizons $a_\pm$ are depicted as functions of  $r$.
A horizontal line on the extended plane at $\sigma=1$ represents a fixed spacetime  $a=$constant. In particular,
$a=0$ corresponds to the Schwarzschild \textbf{BH} spacetime and $a=1$ to the extreme Kerr \textbf{BH}.  In the region $a^2>1$, there are the naked singularity (\textbf{NS}) strips. The angular momentum  of the inner and outer horizons curve are shown explicitly. The momentum $\ell^{\pm}_H=2$ corresponds  to the extreme Kerr \textbf{BH}. On the equatorial plane, the point $r=2$ is the outer ergosurface and the  Schwarzschild \textbf{BH} horizon corresponds to $a=0$. All  quantities are dimensionless.
Right panel:  Light surfaces $r(\ell;a)$ are solutions of the condition
$\mathcal{L}_{\mathcal{N}}=0$ are shown  as functions of the momentum $\ell$ for different \textbf{BH} spin values that are   signed on the curve. Here, $\omega=1/a_0$ corresponds to $r=0$.  $r_g$ is the tangent curve of the metric bundles as function of $\ell$. The radii $r_{\pm}$ denote the outer and inner horizons; the radius $r_\epsilon^+$ is the outer ergoregion on the equatorial plane. All quantities are dimensionless.}\label{Fig:PlotSh}
\end{figure}

\textbf{MBs} also constraint the \emph{stationary observers} as their
     four-velocity  $u^\alpha$ is    a
linear combination of the two Killing vectors $\xi_{(\phi)}$ and $\xi_{(t)}$; therefore,
$
u^\alpha=\gamma\mathcal{L}^{\alpha}= \gamma (\xi_{(t)}^\alpha+\omega \xi_{(\phi)}^\alpha$),  where
 $\gamma$ is a normalization factor and $d\phi/{dt}={u^{\phi}}/{u^t}\equiv\omega$.
The dimensionless quantity $\omega$  is the orbital frequency of the stationary observer  bounded in the range $\omega\in]\omega_-,\omega_+[$. Hence the limiting frequencies  $\omega_{\pm}$  are the photon orbital frequencies solutions of the condition $\mathcal{L_{N}}=0$. $\omega_{\pm}$.
On the other hands
the solutions $r(\omega):\mathcal{L_{N}}=0$ for a fixed $a$ define the light surfaces.
Metric bundles  and light surfaces are represented in Figs\il(\ref{Fig:PlotSh}).

The light-like orbital  frequencies  $\omega_\pm$ and the horizons frequencies $\omega_H^\pm$   satisfy the following relations
\bea&&\label{Eq:card-ian}
\omega_H^\pm\equiv\frac{r_\mp}{2a},\quad\mbox{with}\quad\lim_{r\rightarrow r_\pm}\omega_{\pm}=
\omega_H^\pm,\quad \lim_{r\rightarrow 0}\omega_{\pm}=\pm\frac{1}{a\sqrt{\omega}},\quad\lim_{r\rightarrow\infty}\omega_{\pm}=0,
\eea
where $\omega_H^-\in]+\infty,1/2]$ and   $\omega_H^+\in[1/2,0]$.
The frequency $\omega^{\pm}_H=1/2$ corresponds  to the extreme Kerr \textbf{BH}.

\medskip

\textbf{The extended plane}

A \textbf{MB}  can be represented as  a curve in the  \emph{extended plane}, which is defined as a plane $\mathcal{P}-r$ with $\mathcal{P}$ being a parameter or function that characterizes the Kerr spacetime. In particular, we will use the extended planes
$a-r$ and $\al-r$, where $\al\equiv a \sqrt{\sigma}$ and $r$ is the radial BL coordinate.

\begin{figure}
\centering
\includegraphics[width=9cm]{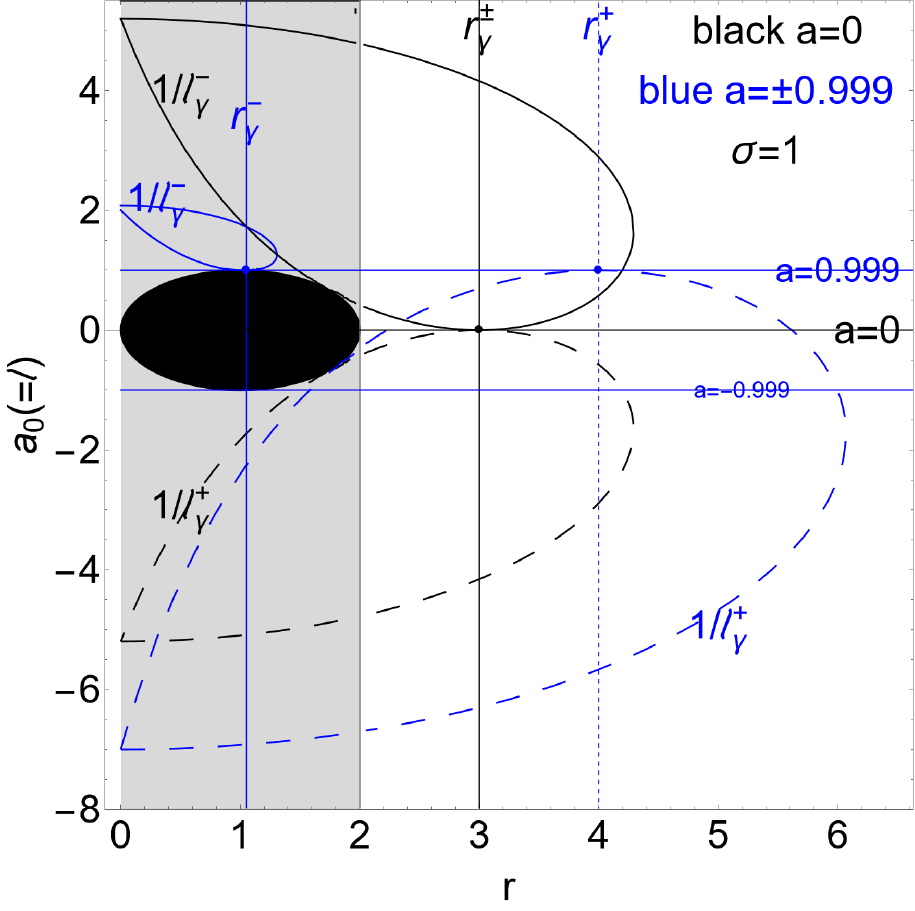}
\includegraphics[width=8.7cm]{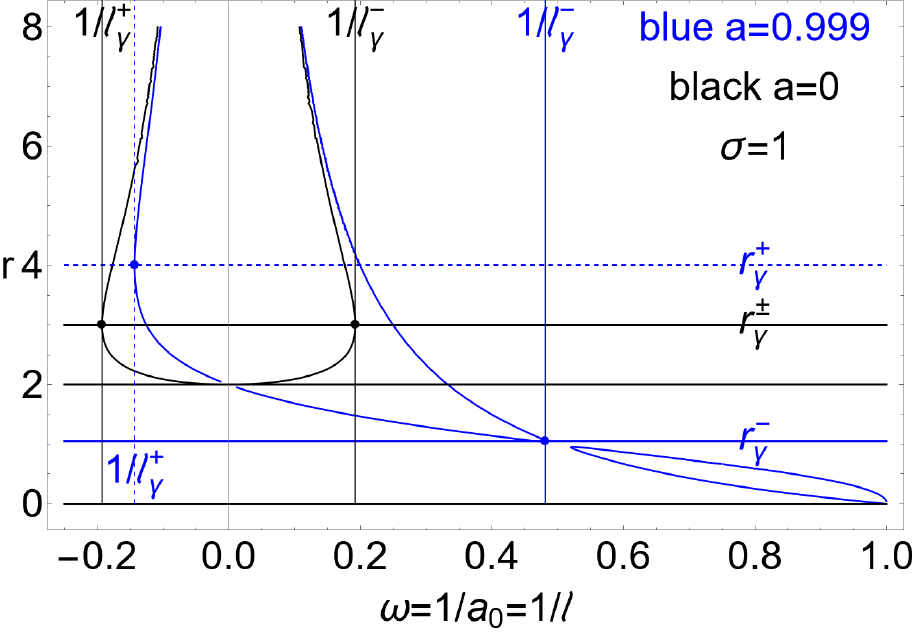}
\caption{Analysis of the equatorial co-rotating (solid curves) and counter-rotating  (dashed curves) photon circular orbits in the extended plane (left panel) and light surfaces (right panel) frame. See the discussion in Sec.\il(\ref{Sec:Metric-bundles}). The case $a=0$, corresponding to the Schwarzschild \textbf{BH} spacetime (black curves), and  the Kerr \textbf{BH} $a=\pm 0.999$ (blue curves) are plotted.  All  quantities are dimensionless.   The extended plane $a-r$  (left panel)  and light surfaces (right panel) $r(\omega)$, as functions of the bundles characteristic frequencies $\omega$ (limiting photon orbital circular  frequencies of the stationary observers), are represented on the equatorial plane $(\sigma=\sin^2\theta=1)$ of the Kerr geometry. On the left panel, the black central region is the \textbf{BH} region bounded by the horizons curves.  The gray region is  the ergoregion $r\in]0,r_\epsilon^+]$, where $r_\epsilon^+=2$ is the outer ergosurface on the equatorial plane.  The metric bundles (\textbf{MBs})  with   characteristic frequency $\omega$ and angular momentum $\ell=1/\omega=a_0=$constant  are shown  for different values of $\ell$,  signed on each curve. The quantity $a_0$ represents the bundle origin spin at $r=0$ (which coincidies  with  the celestial coordinate $\alpha$).
 The radii $r^\pm_{\gamma}$ are the counter-rotating and co-rotating photon circular orbits (marginally circular orbits), respectively. The quantities
$\ell_\gamma^\pm$ are the orbital specific angular momentum  of counter-rotating  ($\ell a<0$, $\omega<0$) and and co-rotating ($\ell a>0$, $\omega>0$) photon circular orbits, respectively, where  $\ell\equiv \mathcal{L}/\Em$  and the constants of motion $ (\mathcal{L},\Em)$ are the angular momentum and energy (as measured at infinity), respectively. The light surfaces $r(\ell;a)$  are solutions of the condition
$\mathcal{L}_{\mathcal{N}}=0$ (on $\sigma=1$) and  are shown  as functions of the momentum $\ell$ for different \textbf{BH} spin values.
 On the extended plane at $\sigma=1$,  a horizontal line  represents a fixed spacetime  $a=$constant.
The points represent the equatorial photons circular orbits (marginally circular orbit) of the spacetime.}\label{Fig:Plotfeatuni}
\end{figure}

 In  the extended plane $a-r$, all the  curves representing \textbf{MBs} (\emph{bundle curves}) are tangent to the horizons curve (the  curve  $a_\pm\equiv \sqrt{r(2-r)}$, which represent   the  Killing horizons of all Kerr \textbf{BHs})--see Figs\il(\ref{Fig:PlotSh})\footnote{This tangency condition implies that  each bundle characteristic frequency $\omega$ coincides with the frequency  $\omega_H^\pm$ of a Killing horizon of the \textbf{BH} tangent to the curve.  All the  metric bundles curves  are  always tangent to the horizon curve  in the extended plane--Figs.(\ref{Fig:Plotfeatuni})-- consequently there is always  a \textbf{BH} horizon $(a,r_\pm)$ which is part of the  bundle, therefore the  bundle characteristic frequencies  are always a \textbf{BH} horizon frequency  (the frequency of the \textbf{BH} horizon correspondent to the point  tangent to the bundle curve in the extended plane).  Hence one could  say that all points $r$ of the  bundle curve  (also in the  \textbf{NSs} spacetimes) are replicas of the \textbf{BH} horizon of  the tangent point.
In this work we constrain  the co-rotating and  counter-rotating replicas in the same spacetime of the tangent point (horizontal lines of Figs\il(\ref{Fig:Plotfeatuni})).}. It follows that  the horizons in the extended plane emerge as the envelope surface of all the metric bundles.

 The concept of metric bundles   and some of their  main features are illustrated  in Fig.\il(\ref{Fig:PlotSh}), in   the extended plane  $a-r$ of the equatorial plane ($\sigma=1$) of the  Kerr  geometry.
 Note that  the extended plane can be extended to  values $a<0$ to consider the counter-rotating photons.

The left panel of Fig.\il(\ref{Fig:PlotSh})  shows  metric bundles at $\sigma=1$   for different   characteristic frequencies $\omega=1/a_0=$constant, where $a_0$ is the  \emph{bundle origin spin} (or bundle origin), defined in Eqs.\il(\ref{Eq:card-ian}).
The right panel of Fig.\il(\ref{Fig:PlotSh}) shows the light surfaces as functions of the bundle origin on the equatorial plane.

The characteristic bundle angular momentum is $\ell\equiv 1/\omega=a_0$.  For $\sigma\in]0,1[$ the bundle origin spin is $\al_0\equiv 1/\omega\sqrt{\sigma}$, corresponding to the solution $a:\mathcal{L}_{\mathcal{N}}=0$ for  $r=0$.

\begin{figure}
\centering
\includegraphics[width=8cm]{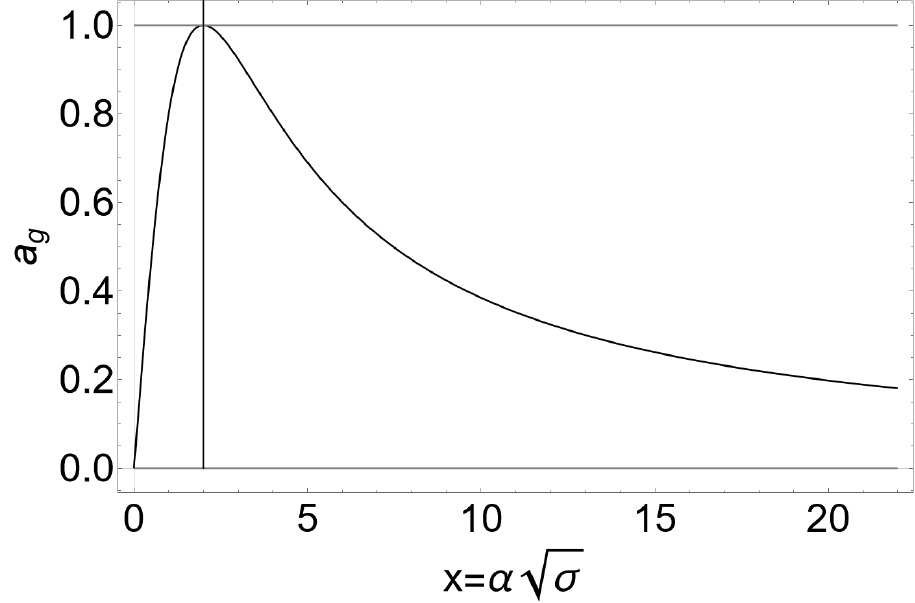}
\includegraphics[width=8cm]{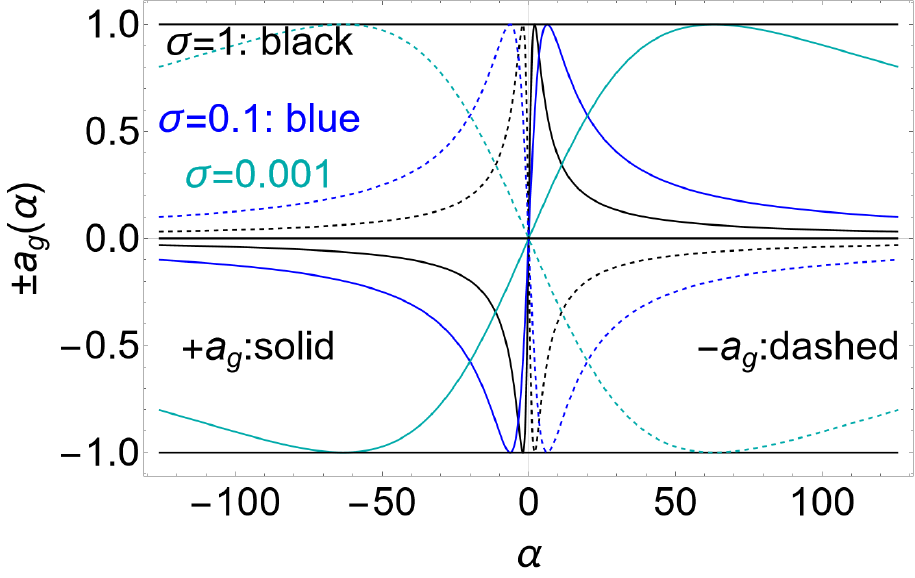}
\caption{Left panel: Tangent spin $a_g$ (tangent spin of the metric bundle curve in the extended plane) as a function of $x=\alpha\sqrt{\sigma}$, equivalent to $a_g(\alpha)$ for $\sigma=1$, where $\alpha=-\ell/\sin\theta$ is the celestial coordinate, which coincides with the bundle origin.  $\ell$ is the photon specific angular momentum and
$\sigma\equiv \sin^2\theta\in[0,1]$.
 Right panel: The functions $\pm a_g(\alpha)$ for different planes $\sigma$. All quantities are dimensionless.}\label{Fig:Plotdatiagx}
\end{figure}

On the extended plane $a-r$, the point $(a_g, r_g)$ is the tangent point of the bundle to the horizons curves.
Therefore, the {bundle origin spin} $\al_0$  bundle tangent radius $r_g(\omega)$ as function of the \textbf{MB} frequency and   bundle tangent spin $a_g(a_0)$  as function of the origin spin $a_0$ are
\bea
&&\label{Eq:bab-lov-what1}
 \al_0(\omega,\sigma)\equiv\pm \frac{\csc (\theta )}{\omega}=\pm \frac{1}{\omega\sqrt{\sigma }}=\frac{\ell}{\sqrt{\sigma}}=-\alpha,  \quad r_g(\omega)=\frac{2}{4 \omega ^2+1},\quad a_g(a_0)\equiv\frac{4 a_0\sqrt{\sigma }}{a_0^2 \sigma +4}.
\eea
there is  $r_g\in[0,1]$ for $r_g=r_-$ and
$r_g\in[1,2]$ for $r_g=r_+$--see for example \cite{bundle-EPJC-complete}.
 Note that $\alpha$  in Eq.\il(\ref{Eq:alpha-def}) coincides with the bundles origin spin $\alpha=-\al_0\equiv -1/(\omega\sqrt{\sigma})\equiv -\ell/\sqrt{\sigma}$ for co-rotating $(\ell>0)$ and counter-rotating photons $(\ell<0)$.
As $\alpha=a_0\in [-\infty,+\infty]$,  then  $a_g(a_0)\in[0,1]$ is a way to express the horizon curves in the extended plane as function of the bundles origin (here restricted to the positive plane $a_g>0$).
In this case, as $\alpha=a_0=1/\omega=\ell\in [-\infty,+\infty]$, $a_g(\alpha)\in [0,1]$ is a way to express the horizons curves in the extended plane as function of the celestial coordinate $\alpha$--see Figs\il(\ref{Fig:Plotdatiagx}).
The function $a_g(a_0)$ establishes a relation between the  bundle tangent spin,  a \textbf{BH} horizon, and a bundle origin spin.
The tangent spin can be associated to two tangent radii, corresponding to an inner or an outer
 horizon (depending on the values of $a_0$).
The function $a_g(\alpha)$ associates  the spin $a_g$ to the celestial coordinate $\alpha$ for a fixed plane $\sigma$ and  can be parameterized for $x=\alpha \sqrt{\sigma}=1/\omega=\ell$.
 Therefore, for a fixed $a=a_g$
 there are two couples of  $(\alpha,\sigma)$ in the \textbf{MBs} framework, corresponding to the inner and outer  horizon frequencies,  respectively, see also Figs\il(\ref{Fig:Plotdatiagx}).

For $\ell=\alpha=2$ there is one couple, which corresponds  to the extreme Kerr \textbf{BH} with  $a_g=1$. Obviously, the product $\alpha \sqrt{\omega}$ is, in general, different for the two couples  because  they correspond  to the angular momentum of the inner and outer horizons.
The maximum point of  $a_g(\alpha)$
for  $x$ or $\alpha$ ($\sigma$) is for
$a_g=1$ and
$\alpha=2/\sqrt{\sigma}$.

From the definition of $\alpha$ we find the spin function
\bea
a=a_g(\alpha)= \frac{4 \alpha  \sqrt{\sigma }}{\alpha ^2 \sigma +4}\in [0,1] ,
\eea
(which coincides with \emph{bundle tangent spin} of  Eq.\il(\ref{Eq:bab-lov-what1})  expressing the \textbf{BH}  horizons curve).
For  $\alpha=0$  and  $\alpha=\infty$ there is    $a_g=0$--Figs\il(\ref{Fig:Plotdatiagx})--left panel.

 In Sec.\il(\ref{Sec:photon}), we discuss in details the relation between  bundles null-orbits and photon circular orbits on the equatorial plane.

\medskip

\textbf{Horizons replicas}

In our analysis we study solutions  of the set $(\mathfrak{R})$ of Eq.\il(\ref{Eq:radial-condition})   with the constraint $\ell=\pm\ell_H^\pm$  therefore on the inner and outer horizons replicas, on the same \textbf{BH} spacetime of the bundle curve tangent point to the horizons,  which are special points on the \textbf{MBs} curves (determined by the horizontal lines in the extended plane).

\medskip

\begin{figure}[h]
\centering
\includegraphics[width=8cm]{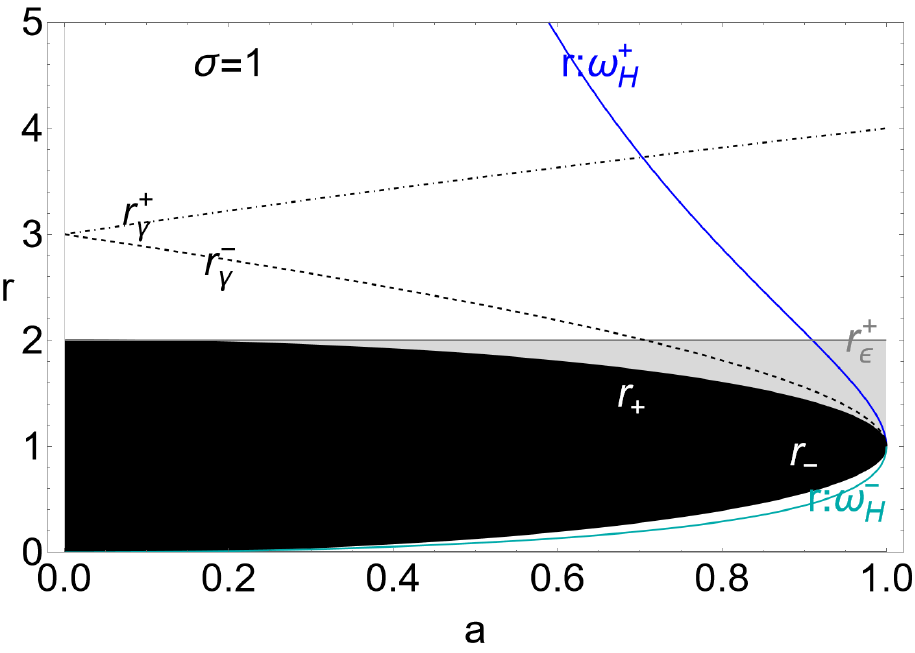}
\caption{Replicas of the inner and outer horizons on the equatorial plane $\sigma\equiv \sin^2\theta=1$. The radii $r_\pm$ are the outer and inner Kerr \textbf{BH} horizons, respectively.  $r_{\epsilon}^+=2$ is the radius of the outer ergoregion on the equatorial plane, $a$ is the \textbf{BH} dimensionless spin, $\omega_H^\pm$ are the outer and inner \textbf{BH} horizon frequencies, respectively. All quantities are dimensionless. Black regions denote \textbf{BH} region. On the equatorial plane, the inner horizon frequencies are confined and the photons cannot reach the observer.  Radii $r_\gamma^\pm$ are the counter-rotating and co-rotating photon circular orbits (marginally circular orbit) on the equatorial plane and  boundary of the Kerr \textbf{BH} spacetime photon sphere.}.\label{Fig:Plotmescol}
\end{figure}

\emph{Horizon replicas} are special light-like circular orbits  of a Kerr spacetime with frequency equal to the \textbf{BH} (inner or outer) horizon frequency, which coincides with  the bundle characteristic frequency in the corresponding   point on the extended plane\footnote{Viceversa, the \textbf{MBs} \emph{horizon confinement}  is a concept   interpreted as due to  the presence of a  ``local  causal ball" in the extended plane, which is   a region, where the \textbf{MBs} are  entirely confined,  i.e.,  there are no horizon replicas  in any other   region of the extended plane outside the causal ball.
In the Kerr extended plane, a causal ball  is  the region upper bounded by a portion of the   inner horizon, which means that the horizon frequencies defined for these points of the inner horizons cannot be measured (locally)  outside this region.}.
	According to the definition of metric bundles, there are clearly   replicas in different  geometries. For instance, to the   points $(\mathcal{P},r)$ and $(\bar{\mathcal{P}},\bar{r})$ on the bundle curve of the extended plane, there can correspond the same light-like  orbital  frequency\footnote{Replicas can, therefore,  be used also  to relate \emph{different} spacetimes. This possibility could be useful in some scenarios featuring transitions from \textbf{BH} to \textbf{BH}, following  the  changing of  some parameters such as charge, mass, and momentum as regulated by the laws of \textbf{BH} thermodynamics, or to explore some theoretical concepts concerning relations between black holes and naked singularities. These aspects have  been explored,  for example, in \cite{Pugliese:2022xry,remnants,Pugliese:2022vni}. However, in this work, we do not consider this situation,    but we fix the \textbf{BH} spin (for any value  $a\in[0,1]$)  and analyze its horizon  replicas.
We do not compare  orbits and horizons of different Kerr solutions in the parameter space but focus instead  on the possibility of detecting  the \textbf{BH} horizons replicas imprint  on its shadows profiles, fixing therefore the  \textbf{BH} spacetime and considering the existence of replicas in the \emph{same} spacetime.}.
%
%
Metric bundles  have been defined by using the  condition that the quantity  $\omega=u^\phi/u^t$ is a constant on photon circular orbits. An analogue definition can be set in terms of $\ell$, which
is a constant of the geodesic motion that is related by $\ell=1/\omega\sqrt{\sigma}$ (with an analogue definition for the counter-rotating motion) with the characteristic frequency $\omega$ of a metric bundle.
Then replica of the outer and inner horizon are
 \bea \label{Eq:Eqmescol}
 \omega_H^\pm: r=\frac{a^2\left(\sqrt{10 r_\mp-a^2}-r_\mp\right)}{2 \left(2r_\mp-a^2\right)}
\eea
%
respectively where $\omega_H^\pm=\omega(\ell_H^\pm,r)$,  are the  outer and inner  horizon frequencies--
see Figs\il(\ref{Fig:Plotmescol}).
We restricted our analysis to the co-rotating case. Furthermore, in certain cases, for example in Kerr naked singularities, the condition $\omega<0$ does not always correspond to  $\ell<0$.
\subsection{On photons orbits   and replicas}\label{Sec:photon}
In this section we focus on the relation between  replicas,  metric bundles,  photons circular orbits, photons sphere (or photon shell) --see also \cite{Perlick,Johnson}.
In Sec.\il(\ref{Sec:photon-shere}) we discuss  the relation between  concepts of  photons spheres   and replicas. In Sec.\il(\ref{Sec:photonu}) we relate circular photons orbits and replicas.

Only some replicas are orbits of the  photon sphere, and only some  photon circular orbits are replicas,  on the other hand only some replicas are solutions of the  system  $(\mathfrak{R})$.
{These particular solutions of   $(\mathfrak{R})$, providing the  shadow boundary (for example black curves of Figs\il(\ref{Fig:Plotzing1}),   are found by solving  $(\mathfrak{R})$  with the constraints $\ell=\pm\ell_H^\pm(a)$, and emerge as points on the  shadow profile (for example vertical lines\footnote{{Considering for example the case for  $a=0.71$ and $\sigma=0.3$ ($\theta=0.5796$), represented in Figs\il(\ref{Fig:Plotzing1}),  for  $\ell_H^-=0.833$ (co-rotating inner horizon replica), there is
$\{\alpha,\beta\}=\{-1.52, 4.5\}$ and
$\{r,q\}=\{2.59, 21.5\}$, and for
$\ell _H^-=-0.833$  (counter--rotating inner horizon replica) there  is
$\{\alpha,\beta\}=\{1.521, 5.036\}$ and
$\{r,q\}=\{2.89988, 26.6343\}$.}} of Figs\il(\ref{Fig:Plotzing1})).}
%
\subsubsection{On photons spheres   and replicas}\label{Sec:photon-shere}
{In general, replicas are defined as solutions of  $\mathcal{L}_{\mathcal{N}}=0$ for  $\omega=$constant or, as  in this work,  for $\ell=$constant.
Here, we look for  special  horizons replicas which are   photon orbits with $\ell=\pm  \ell_H^\pm$ solutions of   $(\mathfrak{R})$. Said differently,  we consider  special  solutions of   $(\mathfrak{R})$, providing the luminous edge of the \textbf{BH} shadow,  constrained by the conditions $\ell=\pm  \ell_H^\pm$. For fixed $a$ and $\sigma$, these solutions are points on the shadow profile.}

Not all the horizons replicas can form the luminous textures of the shadow profile (i.e. they are not solutions of  $(\mathfrak{R})$), and not all the solutions of  $(\mathfrak{R})$ are horizons replicas.
Constraints  for the replicas on the shadow profile are given in details in Sec.\il(\ref{Sec:lblu}). (As discussed on Eq.\il(\ref{Eq:lq-constants}) photon circular  orbit $r_\gamma=3M$ on the Schwarzschild  spacetime is  for example  a solution of  $(\mathfrak{R})$, but it is easy to infer this is not a replica--see  also discussion in Sec.\il(\ref{Sec:photon}).).

\begin{figure}
\centering
\includegraphics[width=9cm]{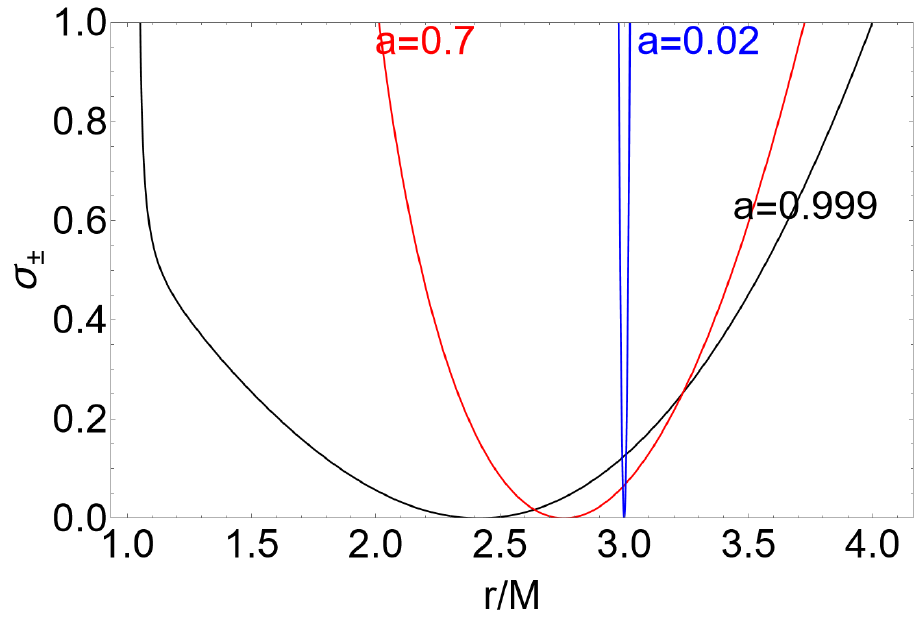}
\includegraphics[width=8.7cm]{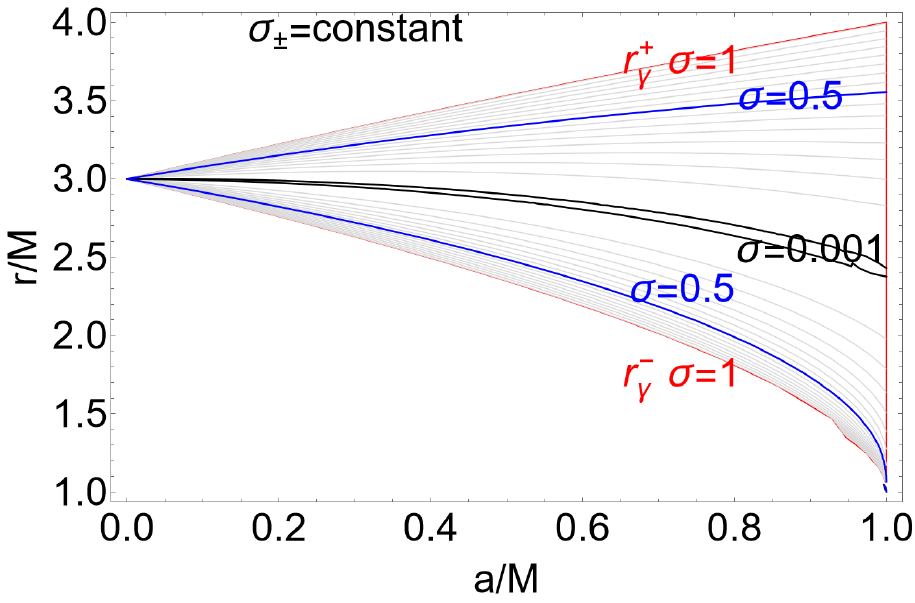}
\caption{Left panel: planes $\sigma_\pm$ of Eq.\il(\ref{Eq:sigmapm}), limiting the Kerr spacetime photon sphere (where $\sigma\equiv \sin^2\theta\in[0,1]$), as function of the radius $r$ for different dimensionless \textbf{BH} spin signed on the curves. Right panel: curves   $\sigma_\pm=$constant  in the plane $r/M-a/M$. Radii $r_\gamma^\pm(a)$,  counter-rotating and co-rotating photon circular orbits respectively on the equatorial plane, and boundary of the photon sphere radius  are also shown (red curves). On $r_\gamma^\pm(a)$ there is $\sigma_\pm=1$.
}\label{Fig:PlotEsis}
\end{figure}

A photon shell or sphere is a region (spherical shell) of  unstable bound photon orbits surrounding the central \textbf{BH}, it is the region
$[r_\gamma^-,r_\gamma^+]$ with $\sigma\in [\sigma_-,\sigma_+]$, where $\sigma_{\pm}$ are functions of $(a,r)$ and $\phi\in[0,2\pi[$, containing bound null geodesics.
The boundary  $r_\gamma^\pm (a)$ are the    counter-rotating and co-rotating circular photon orbits  located on the  equatorial plane  $\theta=\pi/2$ ($\sigma=1$). Their  explicit form is well  known and  can be found for example in \cite{Johnson}.
(In particular for Schwarzschild spacetime, the photon
shell is the two--dimensional sphere  located at  $r_\gamma^\pm = 3M$.)
The limiting angles  are
\bea\label{Eq:sigmapm}
\sigma_\pm\equiv 1-\cos^2\theta_\pm\quad\mbox{where}\quad\quad\cos^2\theta_\pm\equiv\frac{r\left[2  \sqrt{\Delta \left[a^2+r^2 (2 r-3)\right]}- \left(2 a^2+r(r^2-3)\right)\right]}{a^2 (r-1)^2},
\eea
(\cite{Yang,Johnson})-- see  Figs\il(\ref{Fig:PlotEsis})\footnote{We can consider the coordinates $(r,\sigma)$, discussed in Sec.\il(\ref{Sec:lblu}), for replicas solutions of $\mathfrak{(R)}$, with the locations of the photon shells, that is  Figs\il(\ref{Fig:PlotEsis}).}.

The bound geodesics are unstable and after perturbation   can  fall into the \textbf{BH}  or
escape to infinity. Hence,  the
 luminous edge, solutions of $(\mathfrak{R})$ in Eq.\il(\ref{Eq:radial-condition}),  is generated by  nearly bound null geodesics  on the observer screen-- \cite{Bardeen1973}.
All the bound
null geodesics  lie  in the photon sphere. Definition of \textbf{MBs} does not coincide with the  photon  shell, nor contain necessarily orbits of the photons sphere, as explained in details in Sec.\il(\ref{Sec:photon}). First a \textbf{MB} is defined as  the couple   $(a\sqrt{\sigma},r)$, solutions of $\La_\Na=0$ for fixed $\ell$ (hence from normalization condition and the motion symmetries). In the representation of the extended plane  \textbf{MBs} consider  the collection of the spacetimes   (for all  spin $a>0$,  even $a>M$ for \textbf{NSs} spacetimes) where there is a solution of   $\La_\Na=0$ for a fixed value of  $\ell$ (see Figs.(\ref{Fig:Plotfeatuni})).
Not all the solutions of  $\mathcal{L}_\Na=0$ are bound orbits (for example horizons and origin i.e. $r=0$). Some replicas are solutions of  Eq.\il(\ref{Eq:radial-condition}) for certain spins and angles   in accordance with the constraints of  Sec.\il(\ref{Sec:lblu}), resulting in points on the luminous edge of the \textbf{BH} shadow. These special replicas are obtained by  solving   the system $(\mathfrak{R})$ with  $\ell=\pm\ell_H^\pm(a)$.
\subsubsection{On circular photons orbits and replicas}\label{Sec:photonu}
Let us concentrate on the Kerr equatorial plane.  Among the all possible orbits of all bundles on $\theta=\pi/2$, there are  the well-known geodesic photon circular orbits  (which are also the geodesic marginally circular orbits). These orbits are    special bundle orbits (but not necessary   the  \textbf{BH}  horizons replicas of the \textbf{BH} tangent horizon), defined on the special   bundle at $\sigma=1$ having  the characteristic frequency $\omega=1/\ell_\gamma^\pm$, where $\ell_\gamma^\pm\equiv \ell(r_\gamma^\pm)$ is the photon  conserved specific angular momentum $\ell$ given in Eq.\il(\ref{Eq:flo-adding})  of the counter-rotating and co-rotating photon orbits $r_\gamma^\pm$, respectively.  These orbits are  also  special points of the light surfaces bounding the stationary observer orbits.
In Figs\il(\ref{Fig:Plotfeatuni}), we illustrate  in detail the relation between (geodesic) photon circular orbits  on the equatorial plane $r_\gamma^\pm$, metric bundles,  replicas (left panel) and  light surfaces  (right panel).

Figure\il(\ref{Fig:Plotfeatuni})--left panel-- shows the bundles  in the extended plane $a-r$.   Solid curves are the  (equatorial) co-rotating bundles,  dashed curves are the counter-rotating  bundles. Figure\il(\ref{Fig:Plotfeatuni})--right  panel--  shows the light surfaces $r(\omega)$  as functions of the bundles characteristic frequencies $\omega$ (limiting photon circular  frequencies of the stationary observers) on the equatorial plane for different \textbf{BH} spins. The case $a=0$, corresponding to the Schwarzschild \textbf{BH} spacetime,  is represented with black curves, and  the Kerr \textbf{BH}  with $a=\pm 0.999$  with blue curves.   Bundles with   characteristic frequency $\omega$ and angular momentum $\ell=1/\omega=a_0=$constant   ($a_0$ is  the bundle origin spin at $r=0$) are shown  for different values of $\ell$, signed  on each curve.
 The radii $r^\pm_{\gamma}$ are also represented.   The points correspond to  equatorial photon circular orbits (marginally circular orbits) of the spacetime.
 On the extended plane,  a horizontal line  represents a fixed spacetime  $a=$constant.
For $a=0$ , it is $\ell_\gamma^\pm=\mp 3 \sqrt{3}$
on the orbits $r_\gamma^\pm= 3$.
For $a=0.999$, it is
$\ell_\gamma^+=-6.99833$ and
$\ell_\gamma^-=2.07813$, with
$r_\gamma^-=1.05208$ and
$r_\gamma^+=3.99911$, respectively.
It is clear that these orbits  are not horizon replicas.

In this sense, we can read metric bundles  as a generalization of these null (equatorial geodesic) circular orbits, included in the bundles as special points.
  Then marginally circular orbits that exist on the equatorial plane of the Kerr or Schwarzschild spacetimes are, in general, \emph{not} horizon replicas of the Schwarzschild \textbf{BH} and   the Kerr \textbf{BH}  with $a=\pm 0.999$,   respectively. 
\section{Further notes on the shadow analysis}\label{Sec:further-notes}
The functions $
r_H^\pm$, introduced in Sec.\il(\ref{Sec:first-lun})  can be found  as a zero  of the  quantity:
\bea\label{Eq:RHipiu}
&&R_H^+\equiv
a^4+\left(2 a^2+5\right) r^4+4\left(4-a^2\right) r^3+8 a^2+\left(a^4-6 a^2-12\right) r^2+2a^2\left(a^2-4\right) r+r^6-6 r^5.
\eea
Radius $r_H^+$  can be found also  as  solution of 
%
$a(r)=\bar{a}_{H}^+=\tilde{a}_{+}^{-}$, where
\bea
\tilde{a}_{\mp}^\pm\equiv \sqrt{\frac{r [4-(r-3) r (r+1)]\mp 2 \left[\sqrt{(r-1)^3 (r [(r-1) r-4]-4)}\pm 2\right]}{(r+1)^2}}, 
\eea
while $r_H^-$ can be also found  as solution of
%
$a=a_H^-(r)\equiv \tilde{a}_{\mp}^{\pm}$--Figs\il(\ref{Fig:Plotqmminatint}).
Limiting plane $\sigma_\omega^+$, shown in Figs\il(\ref{Fig:Plotqmminatintsigma})--right panel, can be found as a  solution of the equation
\bea&&\label{Eq:sigma-omega-eq}
\sum_{i=0}^{12}n_i \sigma^i=0\quad \mbox{where}
\\\nonumber
&&
n_0\equiv 4096 a^2 \left(a^2-1\right)^2,
\\\nonumber
&& n_1\equiv 2048 (a^2-1) \left(a^6+16 a^4-87 a^2+54\right),
\\\nonumber
&& n_2\equiv 256 \left(a^{10}+220 a^8-466 a^6+972 a^4+1257 a^2-1728\right),
\\\nonumber
&&n_{3}\equiv 512 \left(41 a^{10}+431 a^8-823 a^6-5155 a^4+4754 a^2-1296\right),
\\\nonumber
&&n_4\equiv 256 \left(9 a^{12}+577 a^{10}-5232 a^8+18447 a^6+961 a^4+3606 a^2-1728\right),
\\\nonumber
&& n_5\equiv 256 \left(98 a^{12}-2083 a^{10}+10317 a^8-20461 a^6+6741 a^4+2748 a^2-432\right),
\\\nonumber
&& n_6\equiv 32 a^2 \left(35 a^{12}-2804 a^{10}+19730 a^8-69292 a^6+64139 a^4+51992 a^2+2504\right),
\\\nonumber
&&n_7\equiv -64 a^4 \left(72 a^{10}-1541 a^8+6823 a^6-8371 a^4+14253 a^2-8164\right),
\\\nonumber
&& n_8\equiv 16 a^6 \left(9 a^{10}+417 a^8-2288 a^6+13455 a^4+4481 a^2+566\right),
\\\nonumber
&& n_9\equiv -8 a^8 \left(45 a^8+497 a^6-29 a^4+5507 a^2-3972\right),
\\\nonumber
&&n_{10}\equiv a^{10} \left(a^8+284 a^6+686 a^4+204 a^2-919\right),
\\\nonumber
&&n_{11}\equiv-2 a^{12} \left(a^6+33 a^4-53 a^2+19\right),
\\\nonumber
&&n_{12}\equiv a^{14} \left(a^2-1\right)^2,
\eea
(see also \cite{Pugliese:2021aeb}).
Radius $\bar{r}_H^-(a)$,  discussed in Sec.\il(\ref{Sec:first-lun})m  can be found as a solution of the  which is a zero of the  quantity $\bar{R}_H^-$:
\bea\label{Eq:RHM}
\bar{R}_H^-\equiv a^2+2\left(a^2-3\right) r+\left(a^2+5\right) r^2-4 r^3 +r^4.
\eea
Alternately $\bar{r}_H^-(a)$ can be expressed by   the spin function
  %
$\bar{a}_H^-$: 
 %
 \bea
&&
 \bar{a}_H^-\equiv\sqrt{-\frac{(r-3) r [(r-1) r+2]}{(r+1)^2}}
 \eea
(dotted-dashed curve of Figs\il(\ref{Fig:Plotqmminatint})), where $\bar{r}_H^-(a)$ is a solution of $a=\bar{a}_H^-(r)$.

\end{document}